\documentclass{aa}  
\usepackage{graphicx}
\usepackage{natbib}
\bibpunct{(}{)}{;}{a}{}{,}
\usepackage{subcaption}
\usepackage{float}
\usepackage[symbol]{footmisc}
\usepackage{ulem}
\usepackage{txfonts}
\usepackage[hidelinks]{hyperref}
\hypersetup{
colorlinks=true,
linkcolor=blue,
urlcolor=red,
citecolor=blue
}
\begin{document}

\title{Obscured star clusters in the Inner Milky Way.}
\subtitle{How many massive young clusters are still awaiting 
detection?}

\author{
Akash Gupta\inst{1,2\thanks{Current Institute: 1. Physikalisches
Institut, Universität zu Köln, Zülpicher Str. 77, 50937 Köln}}
\and
Valentin D. Ivanov\inst{2}
\and
Thomas Preibisch\inst{1}
\and
Dante Minniti\inst{3,4}
}

\institute{
Universit\"ats-Sternwarte, Fakult\"at f\"ur Physik, 
Ludwig-Maximilians-Universit\"at M\"unchen, Scheinerstr. 1, 
81679 M\"unchen, Germany\\
\email{agupta@ph1.uni-koeln.de, preibisch@usm.uni-muenchen.de}
\and
European Southern Observatory, Karl-Schwarzschild-Str. 2, 
85748 Garching bei M\"unchen, Germany\\
\email{vivanov@eso.org}
\and
Instituto de Astrof\'isica, Depto. de Ciencias F\'isicas, Facultad 
de Ciencias Exactas, Universidad Andr\'es Bello, Av. Fernandez Concha
700, Las Condes, Santiago, Chile
\and
Vatican Observatory, Vatican City State, V-00120, Italy
}

\date{Received 07/06/2024; accepted 01/11/2024}

\abstract
{The Milky Way star clusters provide important clues for the history
of the star formation in our Galaxy. However, the dust in the disk
and in the innermost regions hides them from the observers.}
{Our goal is twofold. First, to detect new clusters -- we apply the
newest methods for detection of clustering with the best available
wide-field sky surveys in the mid-infrared because they are the least
affected by extinction. Second, we address the question of cluster
detection's completeness, for now limiting it to the most massive
star clusters.}
{This search is based on the mid-infrared Galactic Legacy
Infrared Mid-Plane Survey Extraordinaire (GLIMPSE), to minimize
the effect of dust extinction.
The search Ordering Points To Identify the Clustering Structure (OPTICS)
clustering algorithm is applied to identify clusters,
after excluding the bluest, presumably foreground sources, to
improve the cluster-to-field contrast.
The success rate for cluster identification is estimated with a
semi-empirical simulation that adds clusters, based on the real
objects, to the point source
catalog, to be recovered later with the same search algorithm
that was used in the search for new cluster candidates. As a
first step this is limited to the most massive star clusters
with total mass of $\sim$10$^4$\,M$_\odot$.}
{Our automated search, combined with inspection of the
color-magnitude diagrams and images yielded 659 cluster candidates;
106 of these appear to have been previously identified, suggesting
that a large hidden population of star clusters still exists in the
inner Milky Way. However, the search for the simulated supermassive
clusters achieve a recovery rate of 70-95\,\%, depending on the
distance and extinction toward them.}
{The new candidates -- if confirmed -- indicate that the Milky Way
still harbors a sizeable population of still unknown clusters.
However, they must be objects of modest richness, because our
simulation indicates that there is no substantial  hidden population
of supermassive clusters in the central region of our Galaxy.}


\maketitle
%
\section{Introduction}

Stars are the basic building blocks of galaxies and the vast majority 
of them forms in a clustered environment \citep{2003ARA&A..41...57L}. 
Hence, a complete census of star clusters in the Milky Way is important 
for tracing star formation, chemical enrichment, and galactic structure --
the studies of young massive clusters have an impact on many areas
\citep{2010ARA&A..48..431P}.
A major obstacle in finding star clusters is the extinction that light 
suffers due to the dust along the line of sight and this becomes even 
more severe when the clusters are located closely within the disk of 
our Galaxy. Optical surveys like HIPPARCOS and {\it Gaia} have been 
successful in cataloging the clusters in the solar neighborhood 
\citep[see for example][]{1998AJ....116.2423P, 2017MNRAS.470.2702K} 
but the star clusters farther away -- and behind more dust -- become sou
undetectable in optical surveys. 

Infrared (IR) surveys allow us to find star clusters close to the 
Galactic center, like the massive Arches cluster
\citep{1993ApJ...406..501N, 1995AJ....109.1676N} and to address the
question how close to the center of the Milky Way the clusters can 
survive \citep{2021A&A...648A..86M}. Since then, there 
have been a lot of dedicated IR cluster searches. In a major effort
\cite{2001A&A...376..434D} visually inspected 2MASS \citep[Two Micron 
All Sky Survey;][]{2006AJ....131.1163S} images searching for clusters 
in the Cygnus X region. A wider automated search covering 47\% of the 
sky in the 2MASS point source catalog was carried out by 
\cite{2002A&A...394L...1I}; new clusters were detected by as apparent 
stellar surface density peaks and then verified with a visual 
inspection to avoid artifacts, caused by e.g., dust clouds. Further 
productive  near-IR star cluster searches have been carried out with 
UKIDSS and VVV/VVVX surveys
\citep{2011A&A...532A.131B, vvvcl001, 2012A&A...542A...3S, 2015A&A...581A.120B, 2018MNRAS.481.3902B, 2021A&A...652A.129M}. 

However, the near-IR cluster census is also subjected to 
incompleteness, as shown by \citet[][see their 
figure\,7]{2005A&A...442..195I}, because the extinction in the inner 
Milky Way can be considerable even at these wavelengths \citep[e.g., 
A$_{K_S}$$\sim$2.8\,mag corresponding to A$_{V}$$\sim30$\,mag to some 
clusters;][]{2008A&A...489..583K}. A way to overcome 
this challenge is to turn towards the mid-IR (MIR) wavelengths and 
surveys like GLIMPSE \citep[Galactic Legacy Infrared Mid-Plane 
Survey Extraordinaire;][]{2003PASP..115..953B} and WISE \citep[Wide 
Infrared Survey Explorer;][]{wise_descr}. 

Along this line,  
\cite{2005ApJ...635..560M} used two-dimensional Gaussian fitting to 
the surface density of sources in the GLIMPSE point source catalog 
to identify potential clusters and to estimate their locations and 
sizes. They removed false positives with statistical significance 
tests, finding 59 new, mostly highly embedded clusters. Importantly, 
as we will discuss further, no constraints on the stellar colors 
were imposed. \cite{2013A&A...560A..76M} pointed out that this 
approach would more likely identify nearby bright sources and would 
still miss, despite working in the MIR regimen, many of the reddened 
clusters because of the mix of sources at different distances along 
the line of sight and the ensuing confusion. To counter this 
\cite{2013A&A...560A..76M} only considered red sources with GLIMPSE 
color [4.5]$-$[8.0]$\ge$1\,mag. In this work, no photometric quality 
constraints were applied, which strongly affected the 
contribution of faint stars. Still, their simple automated algorithm 
yielded 75 new clusters, most of them embedded. Finally, 
\cite{2018ApJ...856..152R} identified 923 cluster candidates with 
a visual search in WISE, but the vast majority of them are not 
validated, even through a simple inspection of the 
color-magnitude diagrams (CMDs). 
\cite{2016MNRAS.455.3126C} identified 652 stellar clusters close to 
the galactic plane (mostly embedded) using Radial Density Profile (RDP) 
and CMDs in WISE data raise a good question about how many more 
clusters are present in the disc.

The Milky Way cluster census remains incomplete, despite 
the move towards longer wavelengths and the multiple efforts by 
various teams. In particular, the region close to the Galactic 
center, accessible only through IR observations, is especially 
interesting because of the number of candidate star clusters that 
still need confirmation and determination of physical parameters. 
In this work, we pursue two goals, with corresponding improvements. 
First, we want to address, albeit for now in a limited way, the 
long-neglected question of how complete the existing cluster 
catalogs are. Adding tens or thousands of new clusters means little if
we do not know whether this number of objects amounts to 1 or 50 
or 99\,\% of the entire Milky Way cluster population. Some early 
simulation efforts were carried out by \citet{2010IAUS..266..203I}
and \citet{2010HiA....15..794H} who generated a galaxy-wide
cluster population immersed in dust. Both the cluster and the dust
followed smooth exponential spatial distributions. The model 
calculated the line-of-sight extinction to individual clusters and 
their visibility, estimating what fraction of the Milky Way 
clusters are visible. 
\cite{2005ApJ...635..560M} introduced artificial clusters with 
Gaussian profiles in the GLIMPSE point source catalog only 
reported that all were detected by their search algorithm.
Recently, \citet{2019MNRAS.482.3789N} simulated the cluster 
detection efficiency of the Parzen density estimation method but 
did not tie the results to a particular survey or a type of cluster. 
Parzen Density Estimation, also referred to as Parzen Windows
or Kernel Density Estimation, is a non-parametric method used in 
statistics to estimate the probability density function of a random 
variable based on a finite sample of data points 
\citep{10.1214/aoms/1177704472}.

The most robust way of estimating the detection rate is to carry
out a controlled experiment by adding a sample of simulated star 
clusters. In the most general case the simulated clusters must 
span the full range of cluster parameters -- most importantly 
masses and ages. Another necessary component of such a simulation 
is a detailed 3-dimensional extinction map of our Galaxy. 

Simulating the entire Milky Way cluster population and measuring 
the detection rates for different classes of clusters is difficult
and here, in this pilot study, we only consider the case of the
most massive clusters, analogous to Westerlund\,2
\citep[][]{1961ArA.....2..419W}. It is located relatively nearby at 
4.16$\pm$0.33\, kpc from us and has an age of 1--2\, Myr 
\citep{2015AJ....150...78Z}. It was adopted as an empirical template 
because it falls within the GLIMPSE footprint and a recent
photometric estimate of (3.7$\pm$0.8)$\times$10$^4$M$_\odot$
\citep{2021AJ....161..140Z} places it among the most massive Milky
Way clusters. Only a handful of similar objects are known: 
Arches \citep{1993ApJ...406..501N, 1995AJ....109.1676N}, 
RSGC\,1 \citep{2003A&A...404..223B,2007ApJ...671..781D}, 
Galactic Center Cluster \citep{1968ApJ...151..145B}, 
Quintuplet \citep{1990ApJ...351...89O,1990ApJ...351...83N}, and 
Westerlund\,1\citep{1961PASP...73...51W, 1991AJ....102..642M}. 
Although rare, they are important as examples of galactic building 
blocks and they are local (mini-)analogs of distant star-forming 
galaxies \citep{2010ARA&A..48..431P}.

Finding more massive clusters is challenging, especially in 
the relatively narrow age range when the total cluster 
luminosity in the IR -- where they emit most of their light 
is dominated by pre-main sequence stars, while the rest of
the stars are significantly fainter in the IR. This makes the
question if the Milky Way contains more of them particularly 
relevant.

Our other aim is to complete the Milky Way cluster census further 
while taking advantage of the new and improved search algorithms,
finding undiscovered and highly obscured clusters. The new 
generation 
of algorithms is based on machine learning (ML). To increase our 
chances of cluster identification we consider only sources with 
reliable [3.6] and [4.5] GLIMPSE photometry (errors $\leq$0.2\,mag) 
and only the red stars with [3.6]$-$[4.5]$>$0.6\,mag (see section
\ref{sec:params}). This limit is different than that of
\citet[][\protect{[4.5]$-$[8.0]$>$1.0\,mag}]{2013A&A...560A..76M}, 
who set theirs after \citet{2008AJ....136.2413R}. Our color limit 
is more sensitive to the stars' emission than to the dust because
it omits the [8.0] band and therefore it better removes the stellar
foreground contamination. To underline, our search is optimized for 
the detection of distant and highly reddened clusters -- the type
that are likely to suffer the worst incompleteness.

Our two goals are intertwined because the completeness analysis
requires having at hand a reliable cluster detection tool. Staying
on the side of caution, we consider the identified clusters to be
candidates, because in all fairness we can not exclude false
positives, related to dust clouds and spurious stellar density
variation that can exhibit cluster-like morphology. Deep IR
astrometry or time-consuming follow-up spectroscopy, also in the
IR is needed to confirm the cluster nature of the candidates.

The next section presents the GLIMPSE data, the new clustering 
algorithm, and the properties of the newly found candidates. 
The artificial cluster simulation is described in 
Sec.\,\ref{sec:analysis}, and Sec.\,\ref{sec:conclusions} 
summarizes our results.

\section{Cluster search}

\subsection{The GLIMPSE survey}
We have used the GLIMPSE catalog from 
\cite{2009yCat.2293....0S} which combines GLIMPSE-I v2.0, 
GLIMPSE-II v2.0, and GIMPSE-3D. The catalog is available at 
CDS\footnote{{\href{https://cdsarc.cds.unistra.fr/viz-bin/cat/II/293}{https://cdsarc.cds.unistra.fr/viz-bin/cat/II/293}}} 
and includes:
\begin{itemize}
\item GLIMPSE-I \citep{2003PASP..115..953B} covering the area of 
|$l$|=10-65$^\circ$ and |$b$|$\leq$1$^\circ$, and the Observation 
Validation Strategy region around $l$=240$^\circ$. 
\item GLIMPSE-II \citep{2009yCat.2293....0S} covering the regions 
of:
|$l$|$<$2$^\circ$ and |$b$|$\leq$2$^\circ$,
|$l$|=2$-$5$^\circ$ and |$b$|$\leq$1.5$^\circ$, and 
|$l$|=5$-$10$^\circ$ and |$b$|$\leq$1$^\circ$.
It omits the Galactic center region at 
|$l$|$\leq$1$^\circ$, |$b$|$\leq$0.75$^\circ$, but data for this 
the region from the General Observer program, GALCEN (PID=3677, PI 
Stolovy) is included.
\item GLIMPSE-3D \citep{2009yCat.2293....0S} adds vertical 
extensions stripes of up to |$b$|$<$4.2$^\circ$ at the center of 
the Galaxy and up to |$b$|$<$3$^\circ$ elsewhere.
\end{itemize}
These catalogs overlap in some regions and in such cases, the 
order of priority is GLIMPSE-II, GLIMPSE-I, and GLIMPSE-3D, as 
described in the documentation on the GLIMPSE 
website\footnote{\href{https://irsa.ipac.caltech.edu/data/SPITZER/GLIMPSE/gator_docs/GLIMPSE_colDescriptions.html}{https://irsa.ipac.caltech.edu/data/SPITZER/GLIMPSE/gator\_docs/ GLIMPSE\_colDescriptions.html}}. 
The catalog was cross-matched 
with various other surveys like 2MASS, with a matching radius of 
$0.1\arcsec$, testifying to the excellent astrometric calibration 
of this survey.

\subsection{Search method}

\subsubsection{OPTICS clustering algorithm}

Ordering Points To Identify the Clustering Structure Clustering 
algorithm \citep[OPTICS;][]{10.1145/304181.304187} is a 
density-based clustering algorithm, a further development of 
Density-Based Spatial Clustering of Applications with Noise 
\citep[DBSCAN;][]{10.5555/3001460.3001507}. Unlike DBSCAN that 
combines the objects into clusters, OPTICS orders by their 
reachability-distance to the closest core object from which 
they have been directly density-reachable. 

Here a core object is an object that can be reached by at least 
as many objects, as the required minimal number of members 
(\texttt{min\_samples}) that groups should contain in order 
to be considered a cluster. A cluster grows until the 
reachability-distance to the next potential member exceeds
some limit -- the maximum distance between two stars within 
which they will be considered reachable or members of the 
same cluster (hereafter, \texttt{eps}, measured in degrees). 
The effect these parameters have on the results will be 
discussed in the Sec.\,\ref{sec:params}. 

The method allows for a hierarchy of clusters and given the 
nature of the star formation which tends to occur in structures 
of different sizes, from giant star-forming regions to compact 
star clusters we consider it important to preserve this hierarchy. 
Admittedly, here we do not take full advantage of its power,
optimizing the search parameters to identify distant and 
relatively compact clusters, but this is an important feature if
one aims at the nearer clusters that are embedded in large star
forming regions.

\subsubsection{Data curation and search parameters}\label{sec:params}

It proved unpractical to run the cluster search over the entire 
GLIMPSE footprint, because of the memory and speed requirements to
handle the entire catalog.
Therefore, we fragmented it into 1$^\circ$$\times$1$^\circ$ 
tiles for easier data handling. This size was a compromise between 
computational requirements and the danger of missing clusters that 
are split between two neighboring tiles on the other side. For a 
typical cluster diameter of $\sim$1$\arcmin$, the probability a 
cluster would be split between two tiles is $\sim$3\,\% and we 
ignored these cases.

Next, to ensure good-quality data we considered only GLIMPSE 
sources with photometric errors of $<$0.2\,mag in both [3.6] and 
[4.5] bands. This requirement affects the faint end of the 
apparent luminosity function below [3.6]$\ge$15.5\,mag where the 
color becomes so uncertain that it is impossible to discriminate 
between a typical foreground lower main sequence star and a 
highly reddened distant cluster member star at a level better 
than 3--5 $\sigma$. We also set an upper color limit of 
[3.6]$-$[4.5]=4\,mag, because upon inspection most of reddest 
sources appear to be galaxies. This is not the case for the 
innermost regions of the Milky Way, but our experiments indicated 
that the surface density of these sources is typically too low to 
meet the minimum number of cluster members that we require. 
Therefore, this limit can probably be ignored in further searches.
Furthermore, we have no strict way to confirm the membership of 
individual objects in a cluster, therefore, the members are 
actually suspected or candidate members. 

Last but not least, we applied color criteria as a proxy for the 
reddening and for the distance to individual stars, assuming the 
reddening and the distance are proportional, to zero order. 
Running the search on each color bin separately minimizes the 
field star contamination and improves the cluster-to-field 
contrast.  The extinction and distance are not linearly related
and the clumpy structure of the dust makes this relation even more 
complex. The existing extinction maps, even the 3D ones are of  
little help because they are based on the red clump stars which 
probe the reddening via older populations. An accurate 3D reddening 
map suitable to remove the reddening towards younger clusters will 
not be available until an IR analog of {\it Gaia} becomes available.

To investigate the effect of the color binning we split the 
GLIMPSE sources into multiple [3.6]$-$[4.5] color bins: [0.0, 
0.2], [0.2, 0.4], [0.4, 0.6], [0.6, 4], and run the search on 
each of them separately with the same parameters: requiring a 
minimum of 12 member stars per cluster candidate 
\citep[following classical definitions from: ][]{2010ARA&A..48..431P,1930LicOB..14..154T}
and a maximum radius of the candidate of 6$\arcmin$. Table\,\ref{tab:color_bin} 
shows the number of candidates for the different bins. It is 
lower towards redder bins and many candidates were detected in 
more than one bin: 10094 appear in both the first two bins, 
944 -- in the first three and 66 -- in all four bins. A random 
inspection suggested that a large fraction of candidates are 
spurious, especially in the most populated bluer bins. The 
cluster-to-field contrast increases towards redder bins, because 
of the omitted foreground population, but on the other hand, the 
reddest bins are sparsely populated, and searching in them alone 
would leave too many clusters undetected. In the end, we adopted 
a single color bin of [0.6, 4] as a compromise between two goals: 
to ensure that we identify the reddest and most obscured candidates 
and to exclude the vast blue foreground stellar population.

\begin{table}  
\centering
\caption{Total number of clusters N$_{clusters}$ detected 
by the algorithm in different bins.}
\label{tab:color_bin}
\begin{tabular}{cc}
\hline
[3.6]$-$[4.5] color bin, mag & N$_{clusters}$\\
\hline
$[0.0, 0.2]$ & 49872 \\
$[0.2, 0.4]$ & 18331 \\
$[0.4, 0.6]$ & 4823 \\
$[0.6, 4.0]$ & 2015 \\
\hline
\end{tabular}
\end{table}

\begin{figure}
\centering
\includegraphics[width=8.9cm]{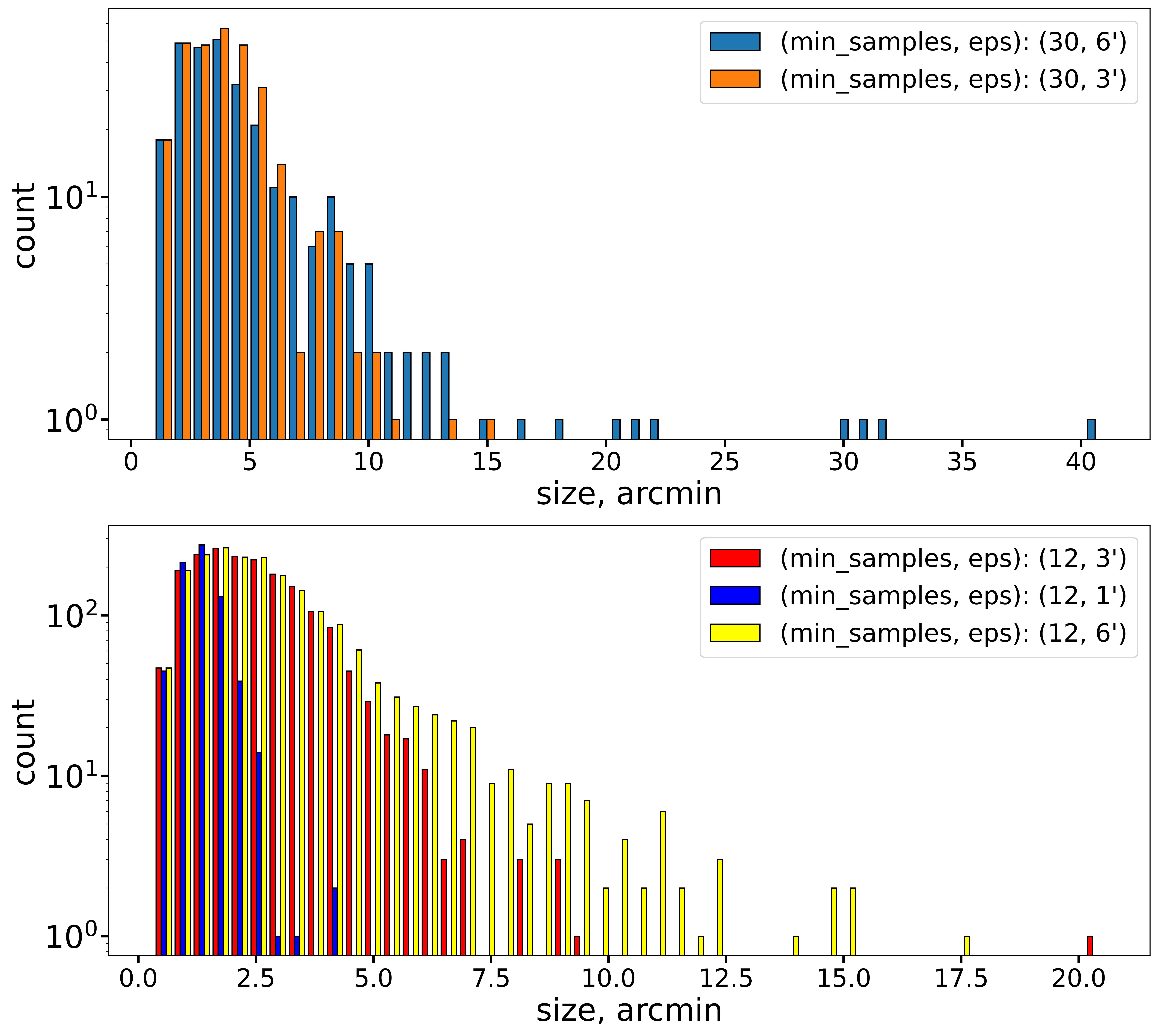}
\caption{Distribution of candidates cluster sizes for different 
search parameters.}
\label{fig:size_histogram}
\end{figure}

We investigated how the OPTICS parameters \texttt{eps} 
\texttt{min\_samples} affect the candidate yield using a 
15$^\circ$$<$$l$$<$35$^\circ$ GLIMPSE cutout. SIMBAD lists 
99 bona fide star clusters in this area. We considered five 
parameter combinations: (\texttt{min\_samples}, \texttt{eps}): 
(30, $6'$), (30, $6'$), (12, $3'$), (12, $1'$) and (12, $6'$).
These values were selected to span the typical cluster sizes 
and richness for known obscured clusters. Our success metric 
was the fraction of the recovered known clusters. The results 
are listed in Table\,\ref{tab:clust_stat}, and the distribution 
of the candidate cluster sizes are shown in 
Fig.\,\ref{fig:size_histogram}. 

This low recovery rate is a concern and we investigated why
so many known clusters remained undetected. One reason was that 
many of them were nearby, subjected to little or negligible 
reddening, and were excluded by our color criterion. Indeed, of 
the 55 undetected clusters, 15 were identified from {\it Gaia} 
observations, 11 are classified in SIMBAD as stellar associations 
and 2 are nearby and NGC clusters, many of which consist of young 
blue stars. This brings our failure rate from $\sim$44\,\% down 
$\sim$27\,\% -- still high and calling for further investigation 
in a forthcoming paper.

A larger \texttt{eps} and a smaller \texttt{min\_samples} 
increase the number of identified candidates; a smaller 
\texttt{eps} and a large \texttt{min\_samples} tend to detect 
more compact candidates, missing sparser clusters. Notably, 
the first two parameter combinations yield few candidates and 
a significantly lower fraction of recovered clusters than the 
other three combinations and these candidates have bigger 
sizes. Therefore, these two combinations are better for 
finding bigger and presumably closer clusters, which fall 
outside of the goal of this work to identify distant and 
heavily obscured clusters, because these types of clusters 
would be readily detected in optical searches, including with 
{\it Gaia}.

The last three parameter combinations require a lower 
number of member stars (12) and a range of minimum radius 
(6$\arcmin$, 3$\arcmin$ and 1$\arcmin$). The first two of
these yield very similar results. Remembering our goal, 
we adopt the ($12, 6'$) parameter combination as a 
compromise -- to be sensitive to compact clusters but also 
to make our detection more complete. Of course, we accept 
the risk of contamination our finding with relatively
nearby and sparse clusters.

Counterintuitively, clusters can have radii greater than 
the maximum distance between two stars {\tt eps}, especially 
in regions of high source density. The {\tt eps} defines the 
maximum distance between density-reachable points that are 
considered neighbors. The algorithm can build a sequence of 
neighbors where the ultimate points are further apart than 
{\tt eps} but are connected with regions of higher density. 
The concept of clusters in OPTICS is broader and the size 
and shape of clusters trace the density distribution of the 
data. Therefore, clusters in OPTICS may span distances 
greater than the adopted maximum distance.

\begin{table}
\centering
\caption{Performance of the OPTICS algorithm for various 
combinations of parameters (\texttt{min\_samples}, \texttt{eps}): 
the number of identified candidates, their median angular radius
in arcmin, the mean number of stars per candidate, the number of
and the fraction of recovered objects with respect to the 99 bona
fide clusters in this region, listed in SIMBAD.}
\label{tab:clust_stat}
\begin{tabular}{@{}c@{ }c@{ }c@{ }c@{ }c@{}}
\hline
 Parame-~& No. of & Median ang. &~Mean no.~& Recovered \\
 ters&~candidates~&~radius,\,arcmin~& of stars &~number (fract.) \\
 \hline
 $(30, 6')$ &  283 & $3.8'$ & 50 & 11 (11.1\%)  \\
 $(30, 3')$ &  288 & $3.8'$ & 51 & 12 (12.1\%)  \\
 $(12, 6')$ & 2014 & $2.4'$ & 19 & 44 (44.4\%)  \\
 $(12, 3')$ & 1853 & $2.3'$ & 19 & 44 (44.4\%)  \\
 $(12, 1')$ &  722 & $1.3'$ & 19 & 33 (33.3\%)  \\
 \hline
\end{tabular}
\end{table} 

In the software implementation of this and the following steps 
we widely used the basic Python modules, AstroPy and Matplotlib
\citep{2018AJ....156..123A,2022ApJ...935..167A,Hunter:2007}.

\subsection{Screening of cluster candidates}\label{sec:screening}

Our search with the adopted parameters yielded 10907 candidates. 
The experience of previous searches has shown that many, if not 
most of them are not real clusters. For example, the clumpy dust 
distribution leads to artificial ``overdensities'' around the 
edges of dark clouds that may significantly exceed the average 
the surface density of the surrounding field, just because the dark 
cloud prevents us from seeing many background sources. This issue 
is only partially alleviated by applying color selection criteria, 
because the color criteria are usually more efficient in removing 
the foreground rather than the background. Furthermore, 
the spatial stellar distribution is uneven itself and can 
produce enhanced surface density regions -- either because of 
large structures, such as spiral arms viewed along the line of 
sights \citep[e.g.,][]{1993MNRAS.261..847K} or because of 
stochastic density peaks \citep[e.g.,][]{2023AJ....165..212A}. 

In the absence of spectroscopic observations, the sole means to 
verify the nature of the candidates is an inspection of the CMDs 
and 3-color images from the available MIR and NIR surveys. The
requirement that the CMDs show some of the typical cluster
sequences adds in effect a degree of physical -- albeit
unquantifiable -- constraints, in addition to the geometric ones,
imposed by the overdensity search.

\begin{figure*}
\centering
\includegraphics[width=9.1cm]{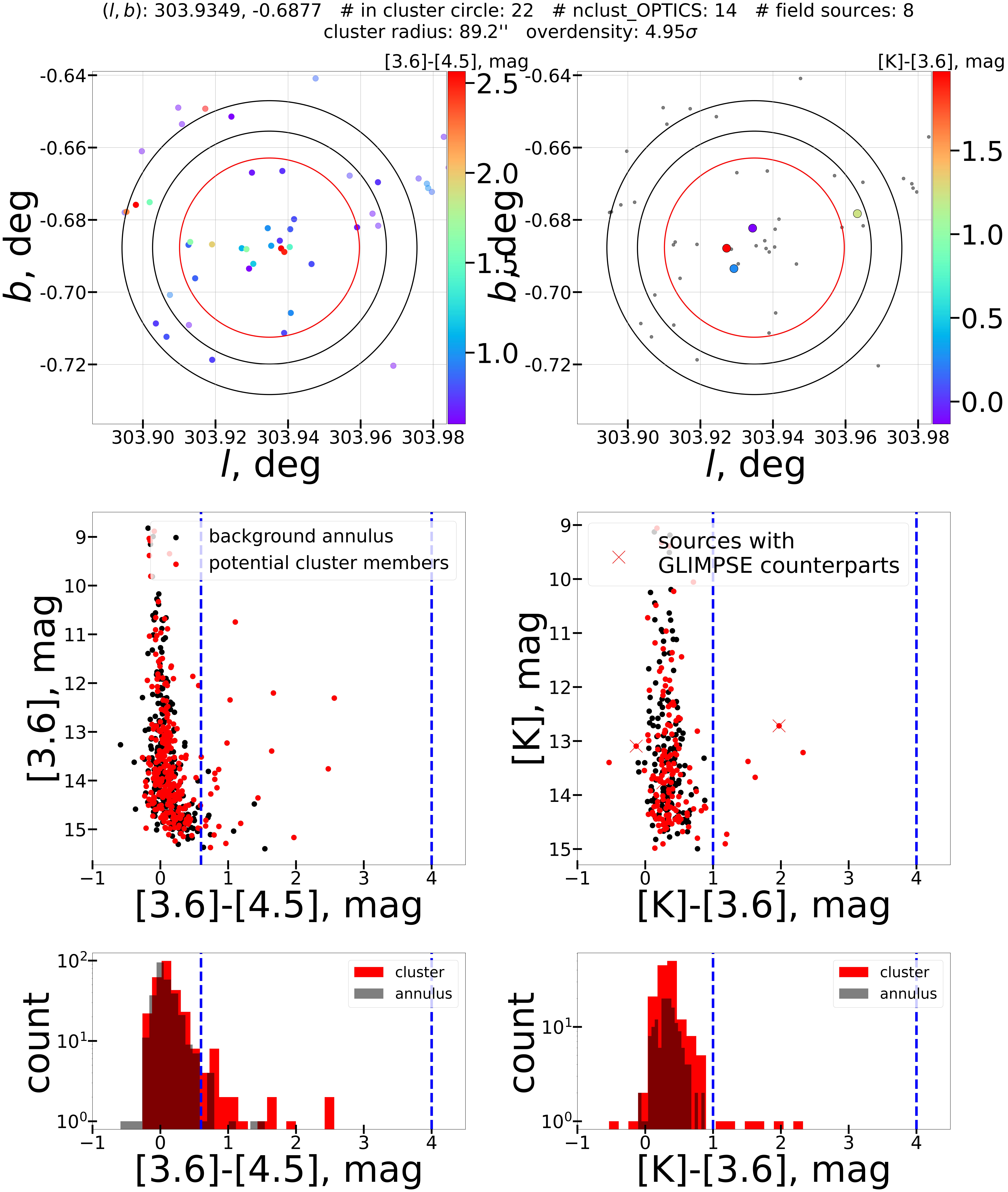}
\includegraphics[width=9.1cm]{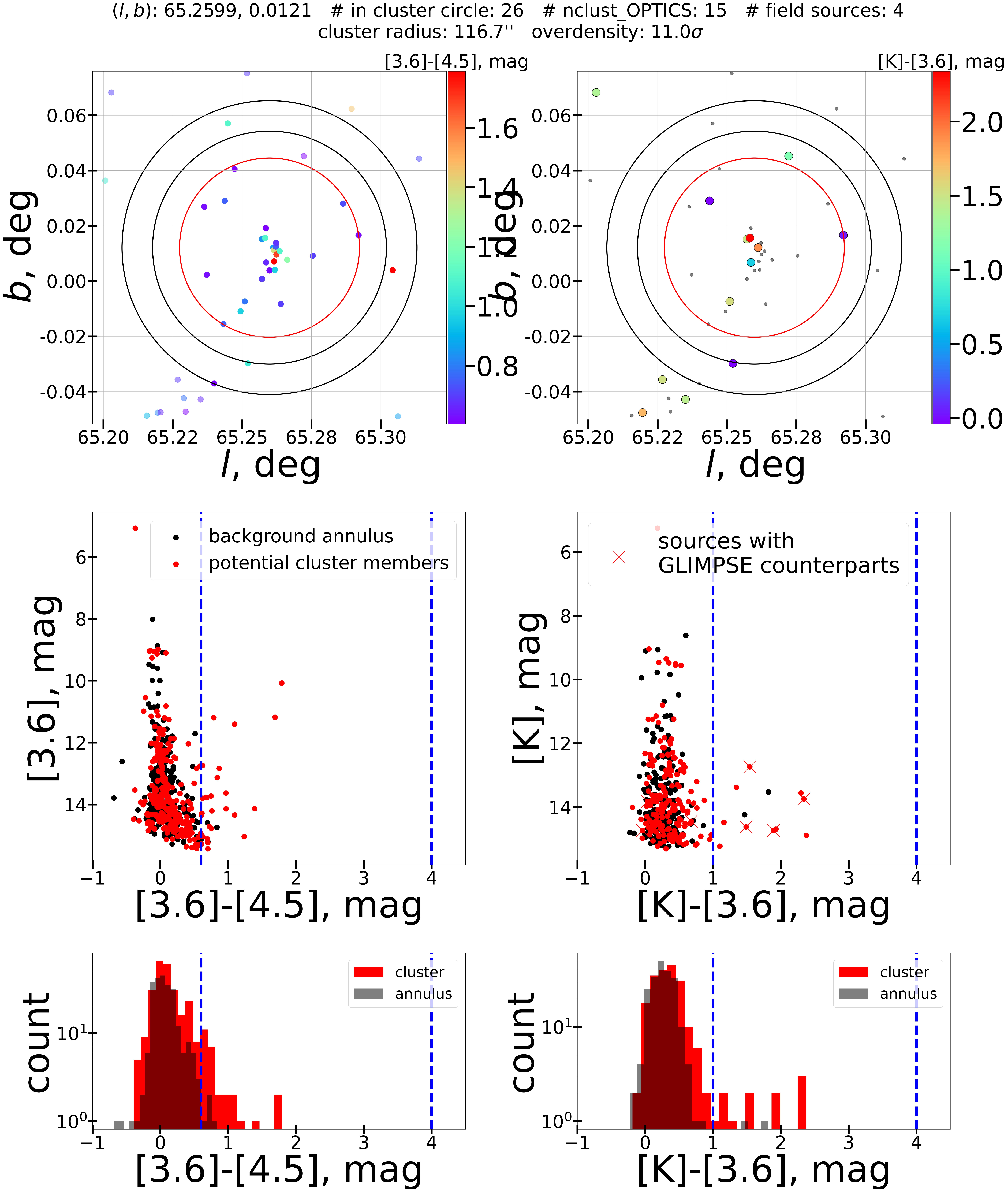}
\caption{Cluster candidate verification plots for a known 
recovered cluster \citep[G3CC 8;][left]{2013A&A...560A..76M}
and for a new candidate (right). The titles list the derived 
cluster parameters. See sections \ref{sec:screening} and
\ref{sec:analysis} for more details. Panels:
{\it Top left} -- mid-IR CMD; red dots are all sources within 
the cluster radius and black dots are all sources in the sky 
annulus.
{\it Top right} -- a map of GLIMPSE sources with colors in 
the bin 0.6\,mag$\le$[3.6]$-$[4.5]$\le$4.0\,mag, color-coded 
by [3.6]$-$[4.5]. The red circle shows the cluster region and 
the black rings -- the sky annulus.
{\it Middle left} -- a combined near/mid-IR CMD; the dots are 
codded lines in the panel above. 
{\it Middle right} -- a map of GLIMPSE--2MASS sources: black 
dots -- all sources, larger dots color-coded by [$K$]$-$[3.6] 
are the object with 0.6\,mag$\le$[3.6]$-$[4.5]$\le$4.0\,mag.
{\it Bottom left} -- Mid-IR [3.6]$-$[4.5] color histogram for
sources in the cluster region (black) and sources in the 
background annulus (red). The potential cluster members cause
the excess at 0.6\,mag$\le$[3.6]$-$[4.5]$\le$4.0\,mag.
{\it Bottom right} -- Near/mid-IR $K$$-$[3.6] color histogram.
The notation is the same as on the previous histogram. Blue 
dashed vertical lines at K$-$[4.5]=1.0\,mag and 4.0\,mag 
are added at to guide the eye only. The potential cluster 
members cause the excess at K$-$[4.5]$\ge$1.0\,mag.
}
\label{fig:primaryscr}
\end{figure*}

\begin{figure}
\centering
\includegraphics[width=2.85cm]{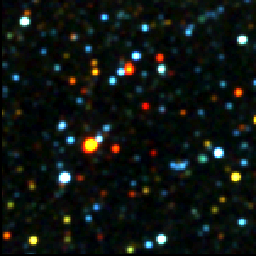}
\includegraphics[width=2.85cm]{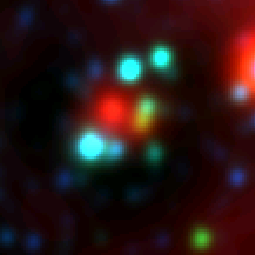}
\includegraphics[width=2.85cm]{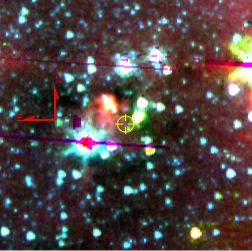}
\caption{Three color images (2.5\arcmin on the side) for the 
cluster in Fig.\,\ref{fig:primaryscr} (left): 2MASS $JHK_S$, 
WISE W1W2W4 and GLIMPSE [3.6][4.5][8.0] (left to right). 
North is Up, East is to the left.} 
\label{fig:example_RGB}
\end{figure}

We calculated several parameters for each candidate:

-- \texttt{($l$,$b$)}: center in the Galactic coordinate system 
determined by taking the median averages of longitudes and 
latitudes of all individual objects grouped by the 
algorithm.

-- \texttt{nclust\_OPTICS}: number of stars ``assigned'' to the 
cluster by the OPTICS clustering algorithm.

-- \texttt{cluster radius}: calculated by taking the mean of 
the Euclidean distance along $l$ and $b$ from the center of the 
two farthest candidate member objects (identified by OPTICS 
algorithm).

-- \texttt{no. cluster members ($\textrm{N}_\textrm{cluster}$)}: 
number of GLIMPSE objects 
(with 0.6\,mag$\le$[3.6]$-$[4.5]$\le$4.0\,mag) that fall inside 
the candidate area is defined as a circle with a diameter equal to 
the derived cluster size.

-- \texttt{no. field sources ($\textrm{N}_\textrm{field}$)}: 
for comparison reasons, we need a sample of fore- and background 
stars in a nearby locus with the same area as that of the cluster. 
This field helps us to verify the overdensity and to estimate 
which parts of the CMD are contaminated by field sources. We 
obtain such a sample within a circular annulus surrounding the 
candidate cluster. The inner radius of the ``sky'' annulus is 
30\,\% larger than the cluster radius to avoid problems due to 
the uncertain cluster size.

-- \texttt{overdensity ($\sigma$)}: excess number of stars in 
the cluster over the number of stars in the field in units of 
background r.m.s. (assuming Poisson statistics): 
\begin{equation}
\sigma = \frac{\textrm{N}_\textrm{cluster}-\textrm{N}_\textrm{field}}{\sqrt{\textrm{N}_\textrm{field}}}
\label{eq:overdensity}
\end{equation}

Note that for the overdensity calculation, we count as members,
all stars within the candidate's locus and all stars in the 
field annulus that meet the same color and error criteria 
adopted for the cluster search. Therefore, the estimated 
overdensity should be treated with caution because it includes 
a certain fraction of fore- and background objects.

\begin{figure*}[ht!]
\centering
\includegraphics[width=\textwidth]{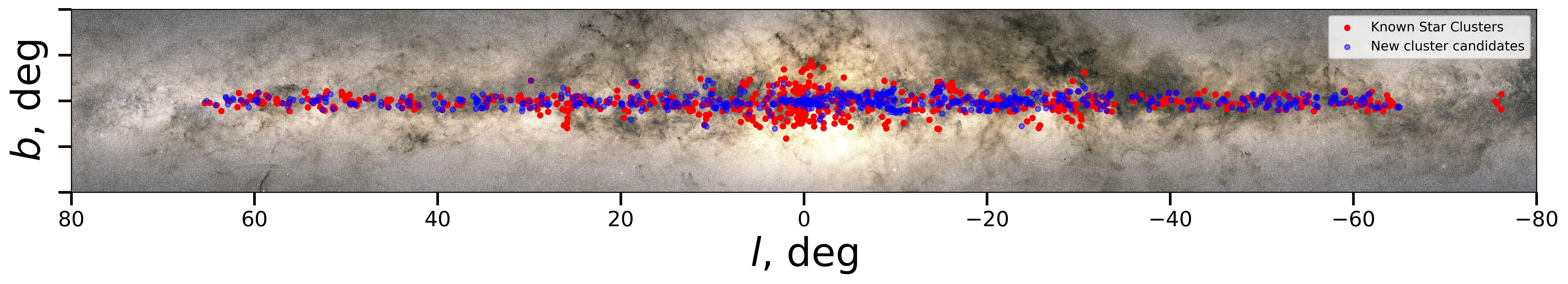}
\caption{Location of newly identified cluster candidates after 
the screening (blue) and the known star clusters (red) in the 
Milky Way.}\label{fig:candidate_distro}
\end{figure*}

To facilitate an efficient screening of thousands of candidates
we created a custom Python at-a-glance visualizer tool that 
combines these numbers and other information we have 
for each object. The main criteria for the true cluster nature 
of a candidate having a statistically significant excess of 
stars near the center, that these stars are more reddened than 
their surrounding counterparts, that they cluster in the CMDs 
in a locus that resembles a reddened main sequence or red giant 
branch and show circular symmetry -- with some leeway for 
asymmetries due to differential reddening, for example. We 
``trained'' on benchmark embedded clusters from 
\citet{2013A&A...560A..76M}. 
Figure\,\ref{fig:primaryscr} shows a CMD inspection image of a 
known embedded cluster (left) and a candidate from our catalog 
(right). The candidate has higher overdensity, supporting the 
clustered nature of this object. This is especially obvious
upon inspection of the mid-IR CMD (upper left panels) if one 
counts the red dots with [3.6]$-$[4.5]$>$0.6\,mag (marked with 
dotted vertical lines) and with [3.6] in the range 8-12\,mag 
and compares their number with the number of the black dots in 
the same locus -- here the red dots mark the stars within the 
cluster region and black dots are the stars in the comparison 
field annulus (both regions have the same areas and are marked 
with red and black circles, respectively, on the right panels).
The inspection 
also included the 3-color 2MASS ($JHK_S$ bands), WISE (W1, W2, 
W4 bands), and GLIMPSE ([3.6], [4.5], and [8.0] bands) images 
of candidates that passed the CMD check. These images for the 
object in Fig.\,\ref{fig:primaryscr} (left) are shown in 
Fig.\,\ref{fig:example_RGB}. The object is virtually invisible 
in the NIR 3-color image. This step is important for excluding 
candidates located next to dense dust clouds that generate a 
necklace-like chain of clusters.

Summarizing, these two steps of screening reduced the sample
size to 659 candidates, listed in Table\,\ref{tab:clusters}.
It shows the adopted nomenclature for their identifier. Their
location on the Milky Way map is shown in
Fig.\,\ref{fig:candidate_distro} (generated with the {\tt python}
module
{\tt mw\_plot}\footnote{\href{https://milkyway-plot.readthedocs.io/en/stable/}{https://milkyway-plot.readthedocs.io/en/stable/}}). 
Importantly, this inspection of the candidates is a subjective 
step that relies heavily on human judgment, despite the 
``training'' on the known clusters.

\begin{table*} 
\caption{Some of the screened cluster candidates. The entire table is available at the CDS.}\label{tab:clusters}
\centering
\begin{tabular}{lrrcccc}
\hline
\multicolumn{1}{c}{Object ID} & \multicolumn{1}{c}{$l$} & \multicolumn{1}{c}{$b$} & overden-~ & ~Radius,~&~No. of~& ~Other \\
\multicolumn{1}{c}{} & \multicolumn{1}{c}{deg} & \multicolumn{1}{c}{deg} & sity, $\sigma$& ~arcmin & stars & ~ID(s) \\
\hline
GIPM 1 & 0.0709 & 0.3743 & 6.00 &  1.8 & 29 \\
GIPM 2 & 0.0797 & $-$0.6430 & 8.49 &  0.9 & 40 \\
GIPM 3 & 0.1323 & $-$0.0621 & 3.27 & 0.9 & 18  \\
GIPM 4 & 0.2507 & 0.0331 & 6.36 & 0.4 & 31 \\
GIPM 5 & 0.2729 & 0.0430 & 9.19 & 0.3 & 44 \\
GIPM 6 & 0.3287 & $-$0.1282 & 5.72 & 0.7 & 28 \\
\multicolumn{7}{c}{...} \\
GIPM 654 & 359.7050 & $-$0.3706 & 6.78 & 2.2 & 32 & [DB2000] 56 \\
GIPM 655 & 359.7471 & $-$0.0992 & 4.99 & 0.7 & 24 \\
GIPM 656 & 359.8198 & $-$0.4394 & 5 & 0.9 & 24 \\
GIPM 657 & 359.8738 & 0.2421 & 6.94 & 0.9 & 33 \\
GIPM 658 & 359.8741 & 0.1393 & 4.62 & 0.9 & 22 \\
GIPM 659 & 359.8778 & $-$0.2187 & 5.00 & 0.6 & 24 \\
\hline
\end{tabular}
\tablefoot{\centering The last column contains other IDs for previously known clusters.}
\end{table*}

\subsection{Properties of the sample of cluster 
candidates}\label{sec:properties}

\begin{figure*}[ht!]
\centering
\includegraphics[width=\linewidth]{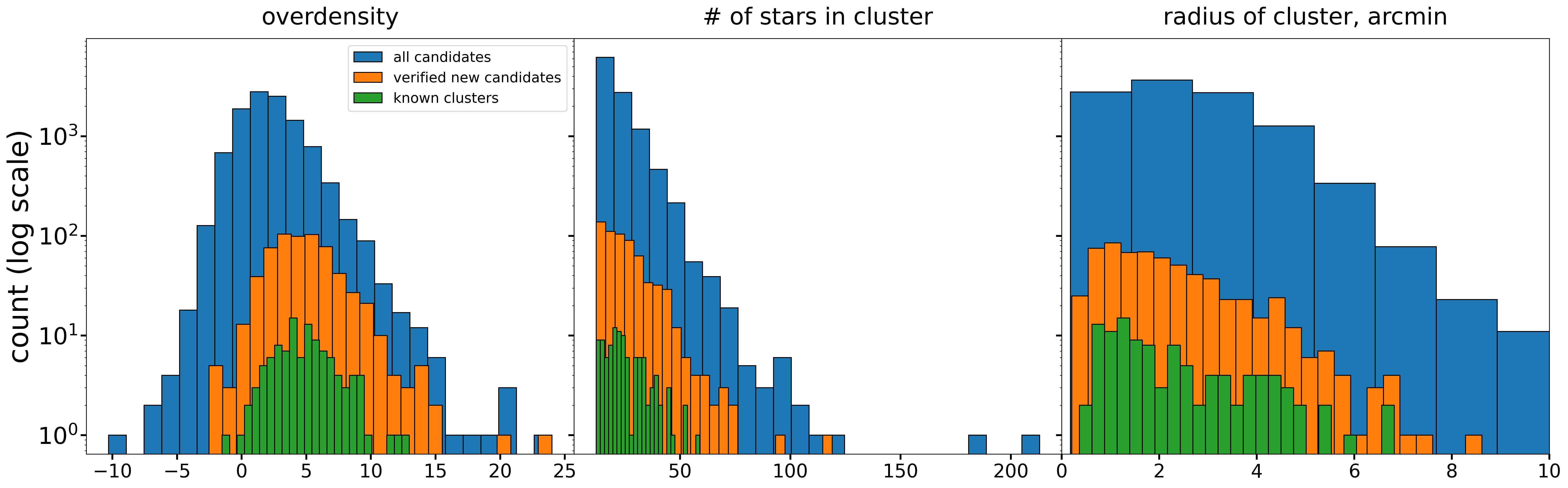}
\caption{Histograms of parameters for all candidates before 
the screening (blue), for remaining candidates after the 
screening (orange), and for the known clusters (green). 
{\it Top left} -- overdensity, {\it top right} -- number of 
member stars, and {\it bottom left} -- 
radius.}\label{fig:hist_attr}
\end{figure*}

A SIMBAD\footnote{\href{http://simbad.cds.unistra.fr/simbad/}{http://simbad.cds.unistra.fr/simbad/}}
search indicated that 106 of the 659 candidates were known:
12 are open clusters, 1 is a globular 
\citep[2MASS GC01;][]{2000AJ....120.1876H} and the rest are 
extremely young embedded star clusters residing in star-forming 
regions. Figure\,\ref{fig:hist_attr} shows histograms of the
derived parameters for all candidates before screening (blue), 
for candidates selected after the screening (orange), and for 
the known clusters (green). 

Typically, the overdensity distributions peak at 4--5$\sigma$
above the fore- and background level. The verified candidates 
and the bonafide clusters tend to present somewhat higher 
overdensities than the average for the initial selection. Some 
overdensities are negative, so there are more stars in the 
annulus than in the cluster region, even for previously known 
clusters. These candidates were still included in our list 
because of the morphology of either their CMD (e.g., an excess 
of redder stars) or of their 3-color images (e.g., showing 
extended emission that is probably associated with dust in the 
mid-IR or with gas in the near-IR). Most of the highest 
overdensities with $\sigma$$\ge$20 tend to exhibit 
cluster-like CMDs and/or morphologies and pass through the 
screening. The rejected high-overdensity candidates
are located at the edges of dark clouds.

The histogram of the number of member stars for the screened 
sample spans a similar range as the histogram of known clusters 
and they both have similar shapes. The few outliers again are 
objects towards the Galactic Center where the crowding is very 
high. Almost all candidates with more than 75 stars from the 
initial sample were rejected.
Finally, the range of measured radii spans 0.5--8.5\arcmin. This 
includes somewhat larger objects than most known clusters, but
among the candidates have radii up to 7\arcmin. Most of the 
largest candidates in the initial sample are rejected. We
underline that these are angular sizes and the actual physical 
sizes of our clusters are unknown because we lack distance 
measurement to each object. Therefore, a nearby and a more 
distant cluster with the same angular size can have a very 
different nature. The color selection that excludes the bluest 
and presumably located nearest to us cluster candidates works 
against this bias, reducing the potential distance-related 
difference.

X-ray emission from the strong coronal activity is known to 
occur in many young stars 
\citep[see e.g.,][]{2007prpl.conf..313F, 2005ApJS..160..401P}. 
Therefore, the presence of X-ray sources would lend some
support to the genuine nature of these candidate clusters, 
which are real young star-forming objects. Prompted by this 
consideration, we cross-matched our list of candidate clusters
with the {\it Chandra}\footnote{\url{https://chandra.harvard.edu/}} 
 Source Catalog \citep{2010ApJS..189...37E} and found that some 
of them indeed contain many candidate members with X-ray 
counterparts: 69 candidates out of 659 have at least one {\it 
Chandra} source within 5\arcsec\ from any of the GLIMPSE 
sources that fall within the cluster radius. This is about
$\sim$19\,\% of 371 candidates that fall within a {\it Chandra} 
pointing. Only some well-known clusters contain multiple X-ray 
sources. The lack of {\it Chandra} counterparts may also be due 
to the low X-ray luminosity of the sources and the heterogeneous 
nature of the archival {\it Chandra} observations that were 
used to build the catalog probably also plays a role.

\subsection{Properties of individual clusters of interest}\label{sec:extremes}

\begin{figure}
\centering
\includegraphics[width=2.85cm]{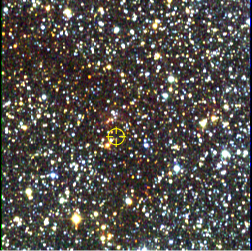}
\includegraphics[width=2.85cm]{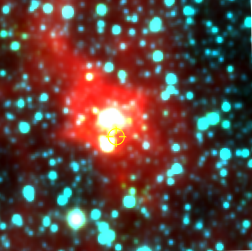}
\includegraphics[width=2.85cm]{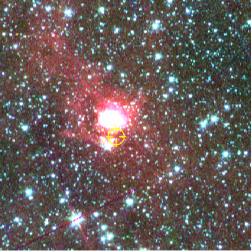}
\includegraphics[width=2.85cm]{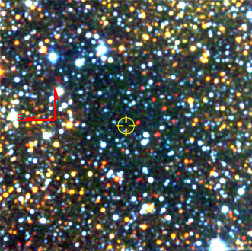}
\includegraphics[width=2.85cm]{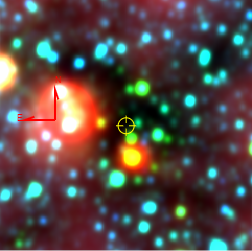}
\includegraphics[width=2.85cm]{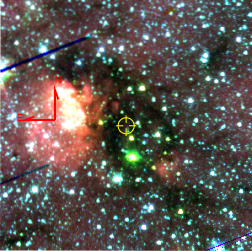}
\includegraphics[width=2.85cm]{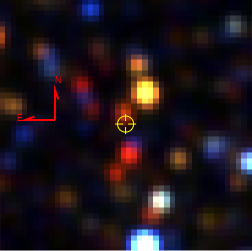}
\includegraphics[width=2.85cm]{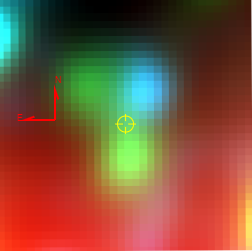}
\includegraphics[width=2.85cm]{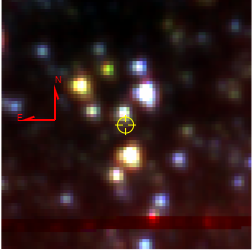}
\caption{Three color 2MASS $JHK_S$, WISE W1W2W4 and GLIMPSE 
[3.6][4.5][8.0] (left to right) images for the most extreme objects 
in our sample: GIPM 257, the largest in 
size (top row; image size 16.6$\times$16.6\arcmin), 
GIPM 402, the richest in number of 
suspected members (middle row; 5.5$\times$5.5\arcmin) and the 
GIPM 8, the smallest in size (bottom 
row; 45$\times$45\arcsec). North is always Up, and East is to the left.} 
\label{fig:extreme}
\end{figure}

Some examples of extreme cluster candidates are shown in 
Fig.\,\ref{fig:extreme}. The first is the largest (top row) 
and it is associated with dust emission in the longest 
wavelength bands. The second is the richest (middle row) and 
it is associated with a dark cloud. The third (bottom row) 
is the smallest in size and it stands out over the field 
stars as a compact group of bright sources. The three of them 
showcase the morphological features that we use to recognize a 
cluster candidate. One typical morphological feature is that 
the suspected 
members are inherently red, which is to be expected as we
are searching for clusters in the inner Galaxy, subjected by 
significant reddening along the line of sight or extremely 
young objects that are still embedded in their parent dust 
clouds. 

\begin{figure*}[ht!]
\centering
\includegraphics[width=0.41\textwidth]{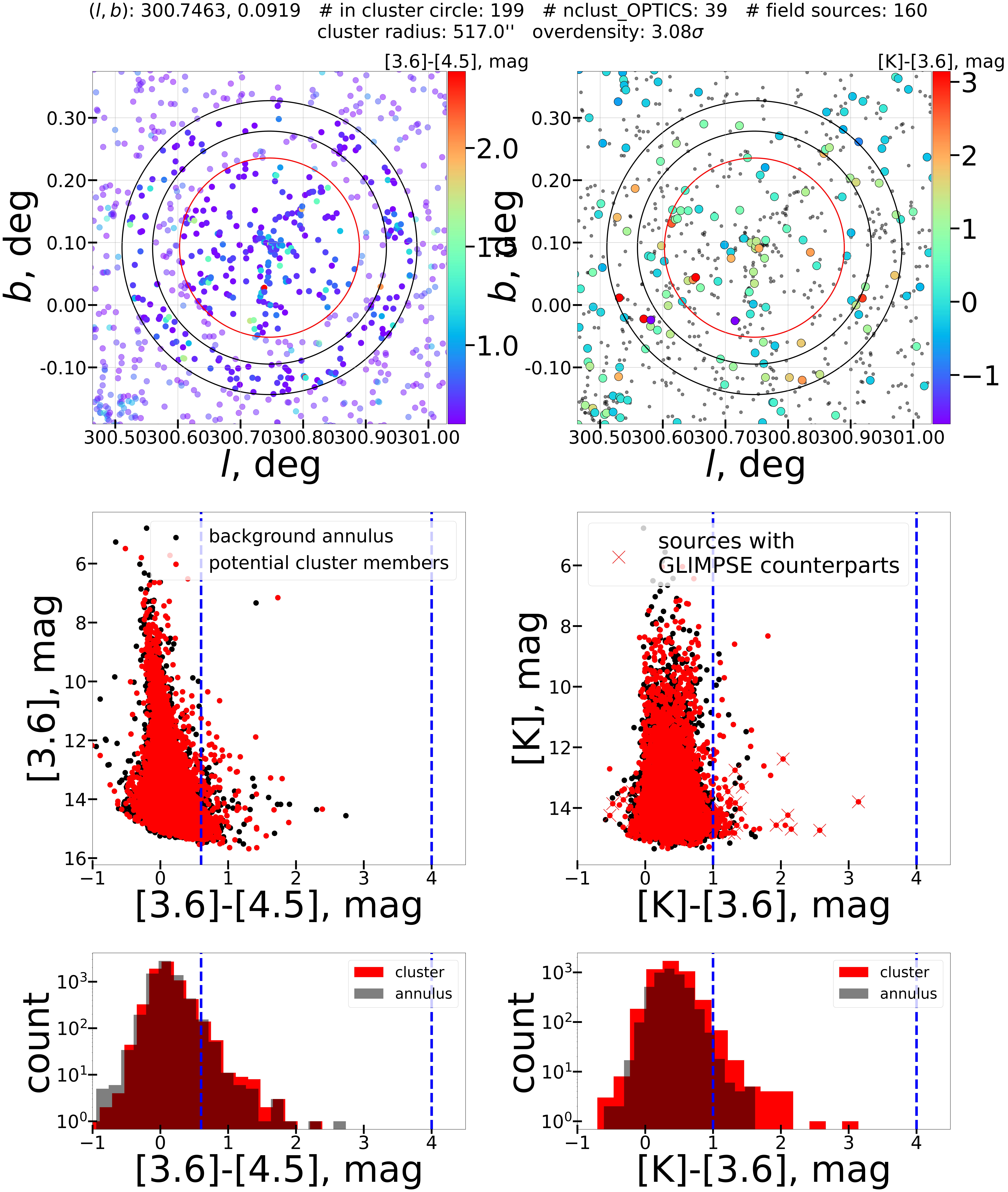}
\includegraphics[width=0.41\textwidth]{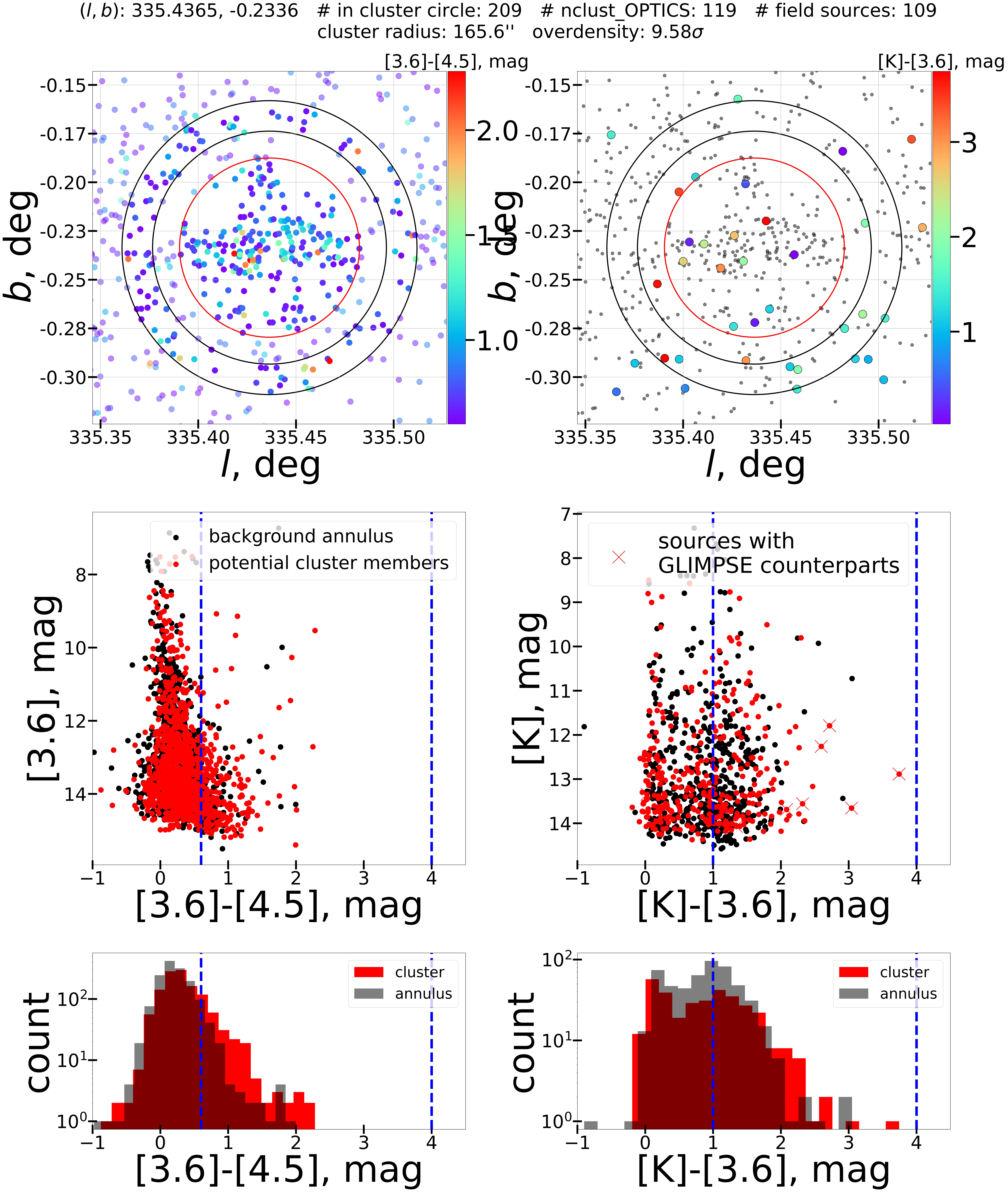}
\includegraphics[width=0.41\linewidth]{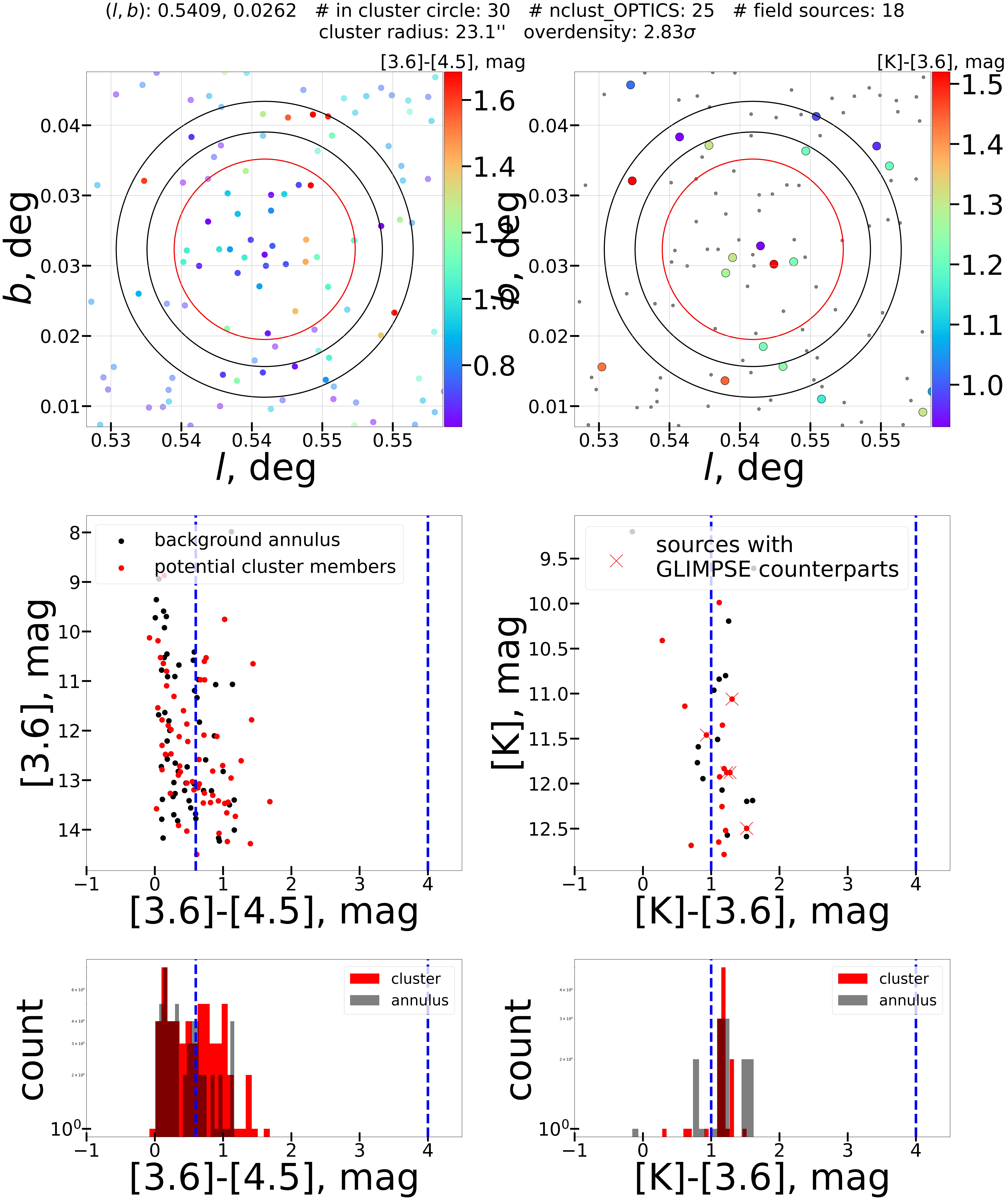}

\caption{Verification plots, similar to Fig.\,\ref{fig:primaryscr}, 
for some of the most extreme candidates. See\,\ref{sec:extremes} 
for discussion of individual objects.\vspace{-0.1cm}}\label{fig:extreme_CMDs}
\end{figure*}

To demonstrate our candidates' range of properties, we discuss 
three of the most extreme ones. Their verification plots, 
including CMDs and maps, are shown in Fig.\,\ref{fig:extreme_CMDs} 
(from top to bottom). 

{\bf GIPM 257 (l,b: 300.7463$^\circ$, +0.0919$^\circ$)} is the 
largest cluster candidate from our search with a preliminary 
radius of 8.62\arcmin. There is an excess of redder sources 
within the cluster region suggesting this is a site of ongoing 
star formation. Indeed, data from the Millimetre Astronomy 
Legacy Team 90 GHz Survey 
\citep[MALT90;][]{2011ApJS..197...25F,2013PASA...30...57J} 
helped to recognize this as a high-mass star-forming region 
with dense cores 
\citep[\protect{[HJF2013] G300.747+00.096};][]{2013ApJ...777..157H}
and young stellar objects (YSOs)
\citep[AGAL G300.748+00.097;][]{2016PASA...33...30R}.


{\bf GIPM 402 (l,b: 335.436468$^\circ$, $-$0.233556$^\circ$)} is 
the richest cluster candidate on our list. It contains the highest 
number of suspected members: 119 as detected by the clustering 
algorithm and 209 within the search radius of the estimated radius 
of the cluster (2.76\arcmin). Previous observations have identified 
many YSOs 
\citep[SSTGLMC G335.4318$-$00.2353 and others;][]{2008AJ....136.2413R}
dense cores
\citep[AGAL G335.441$-$00.237 and others;][]{2013A&A...549A..45C}
and sub-millimeter sources 
\citep[\protect{[LCW2019] GS335.4410$-$00.2323} and others;][]{2019A&A...631A..72L},
consistent with a young cluster. The 3-color near- and mid-IR 
images suggest the presence of a dark cloud.

{\bf GIPM 8 (l,b: 0.5409$^\circ$, +0.0262$^\circ$)} is the densest 
candidate identified by our search, where the density is the number 
of stars divided by cluster area. This is largely because this
object appears compact on the sky with a radius of only 0.4\arcmin. 
Note that this surface density is different from the overdensity in 
units of background r.m.s. (Eq.\,\ref{eq:overdensity}) which is 2.8
for this object. The region is abundant with sub-millimeter
\citep[JCMTSE J174647.7$-$282708 and others;][]{2008ApJS..175..277D,2018ApJS..234...22P}
and X-ray sources \citep[CXOGCS J174646.5$-$282708 and others;][]{2006ApJS..165..173M}. 
The sub-millimeter and X-ray emission may be emitted in dense
cloud cores or originate from coronal activity in young stars
\citep{2018ApJS..234...22P,2007prpl.conf..313F}.
Similarly to the previous candidates, the 3-color near- and mid-IR 
images reveal the presence of dust around the cluster.

\section{Cluster detection completeness}\label{sec:analysis}

\subsection{Generation of artificial clusters}\label{sec:gen_clust}

We adopted an empirical approach to generate artificial 
clusters, instead of the traditional theoretical method that 
starts with a stellar initial mass function and radial 
surface density profiles to produce a fully synthetic object. 
We took the existing cluster Westerlund\,2 with its well-known
parameters as representative of its class cleaned it 
statistically from contaminating field stars, and ``moved'' 
it to a larger distance, accounting for the increased crowding 
and increasing the extinction as appropriate. This method only 
can simulate clusters that are located further away than the 
prototype object, but it is model-independent and free-form 
model-related biases.

First, we removed the field contamination from the sources in
the region of the prototype cluster Westerlund\,2. We adopted a
cluster
radius 1\arcmin, then we defined an annulus (with the same area 
as the cluster region and centered on the cluster) where we 
sample the field population for statistical decontamination of
the cluster. Experiments showed that the exact inner radius of
the annulus is not critical, as long as we keep it within 
1--2\arcmin\ from the cluster region -- the remaining number of 
stars after the decontamination changes at a few percent level. 
We choose to place the annulus next to the cluster region to 
minimize the effect of any large-scale stellar surface density 
variations across the field.

\begin{figure}
\centering
\includegraphics[width=8.5cm]{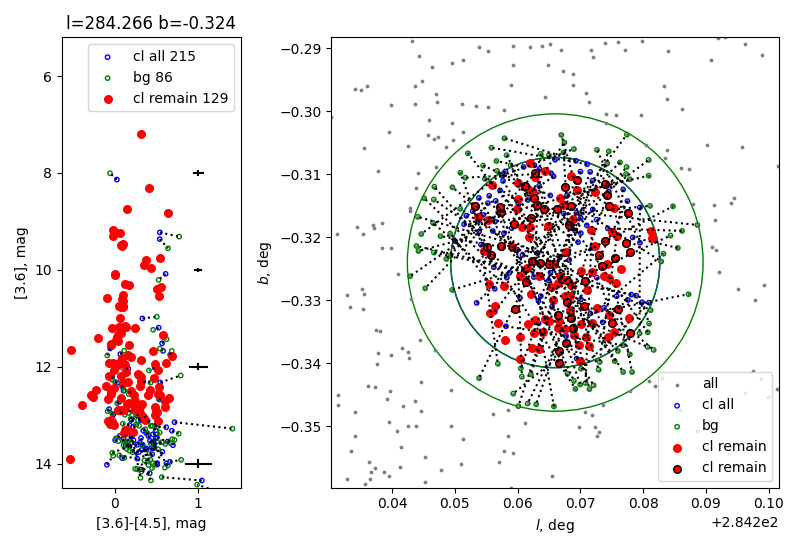}
\includegraphics[width=8.5cm]{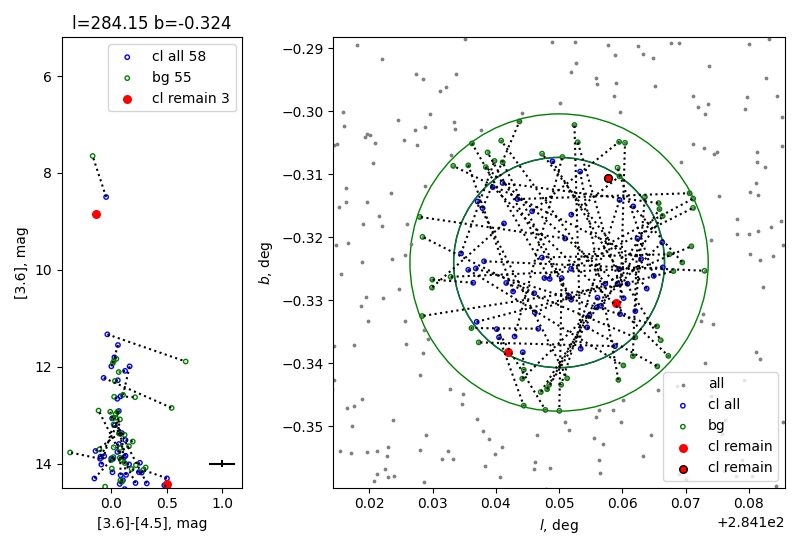}
\caption{Example of statistical decontamination of the cluster 
Westerlund\,2 region (top) and an empty field nearby (bottom),
shown to verify the removal procedure.
The galactic coordinates of the two regions are marked on the
top. The left panels show the GLIMPSE CMDs and the right --
the maps.
The blue circles are all sources within the cluster radius
(adopted 1\arcmin), the green circles -- all sources in an 
adjustment circular annulus (marked with green lines) with the 
same area as the cluster region, and the gray dots are all sources
outside both these two regions. The black dotted lines connect
each removed cluster star with the corresponding field star. 
Solid red dots mark the remaining clusters of stars. The numbers 
in the legend give the number of each type of object.}
\label{fig:decontam}
\end{figure}

The actual decontamination was carried out in the CMD 
space -- for each object in the field annulus we removed the 
closest in color and magnitude to the objects within the 
cluster region, starting from a random source, until one cluster 
object is removed for every object in the field annulus. The 
procedure is illustrated in Fig.\,\ref{fig:decontam} where for 
verification purposes we also show a decontamination of a pure 
field region where nearly all sources in the ``cluster'' region 
have been removed, as expected.

\begin{figure}
\centering
\includegraphics[width=9.4cm]{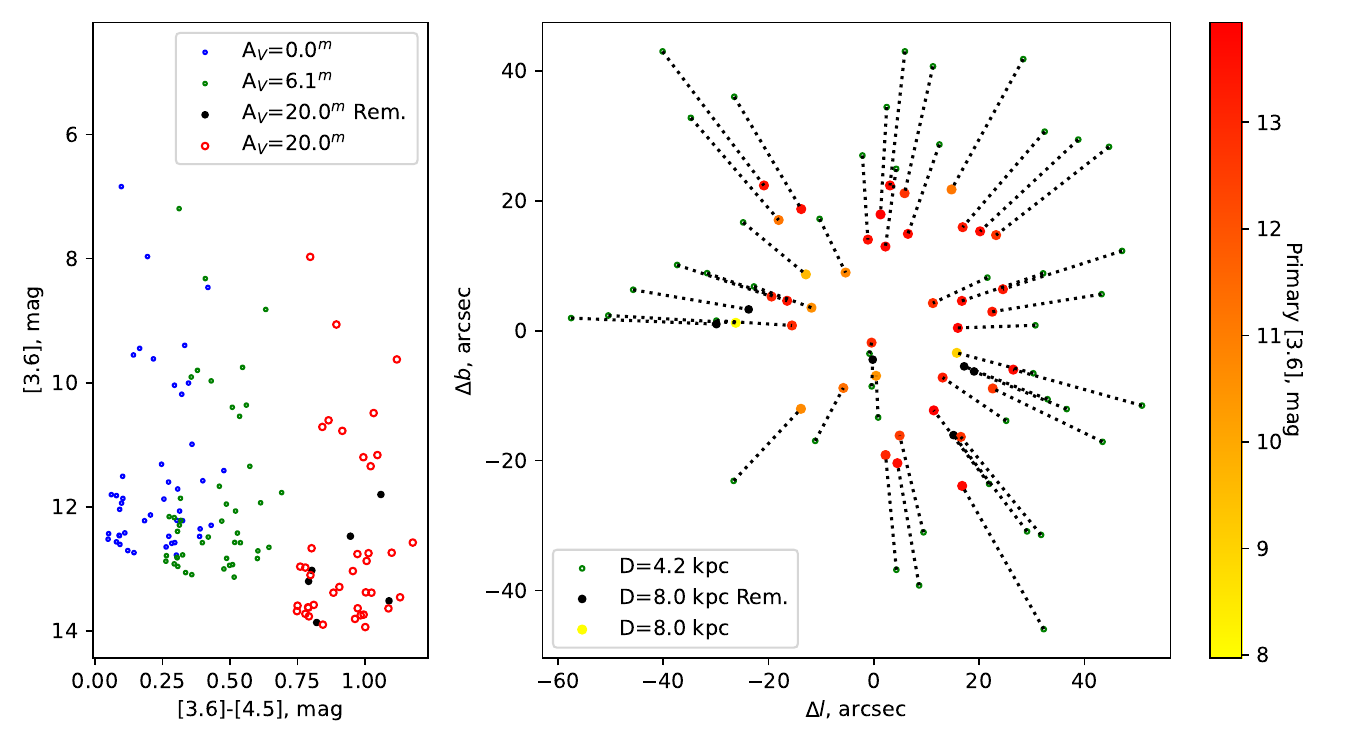}
\caption{Example of constructing an artificial cluster out of 
the ``pure'' Westerlund\,2 population.
The CMD is shown in the left panel. The green points mark the 
apparent position of the member stars as they are in Westerlund\,2,
the blue points indicate correction for reddening and the red 
points reflect the additional reddening to the newly inserted 
artificial cluster. The distance modulus change is not 
applied here. 
The right panel shows the change in the apparent position of 
the member stars as they are ``moved'' from the distance of 
Westerlund\,2 further out to the distance of the new artificial
cluster.
Black points on both panels are stars that are removed or
merged because they came closer than the adopted limit to 
other stars that are brighter than them. For details see 
Sec\,\ref{sec:gen_clust}.}
\label{fig:scaling}
\end{figure}

The next step was to shift the ``pure'' Westerlund\,2 population
to a grid of predefined positions, extinctions, and reddenings
where the artificial clusters would be located. This procedure 
is shown in Fig.\,\ref{fig:scaling}. First, we corrected for the 
distance modulus and the extinction of Westerlund\,2 itself --
this is the move from the green to the blue points on the CMD
(left panel). Then, the data must undergo three modifications: 
adding to the apparent magnitudes the respective distance modulus, 
adding to the color the reddening according to the extinction law 
of \citet{1985ApJ...288..618R} -- this is the move from the blue 
to red points on the CMD, -- and accounting for the decreased 
angular separation between sources, because of the increased 
distance. The latter effect would have merged some nearby sources 
if they were observed with {\it Spitzer} at the newly 
adopted distance. These are marked with black dots on both panels 
and labeled as removed sources, although their flux was 
preserved and merged with nearby sources that are brighter than
each of them. The flux merging uses Pogson's law. 

The condition to merge sources was based on a study of the GLIMPSE 
point source catalog in the inner Milky Way, to determine how 
close the nearest source can be as a function of the ``primary'' 
star's magnitude: for [3.6]$\sim$8\,mag the closest ``secondaries'' 
are usually at least $\sim$4\,arcsec\ away; for [3.6]$\sim$10\,mag 
-- at least $\sim$3\,arcsec\ away and for fainter stars -- at least
$\sim$2\,arcsec\ away. Most likely these are projected, 
not physical binaries. We fitted a linear relation through these 
three points and merger sources that come closer than this limit 
when we move the prototype Westerlund\,2 sources to the position of
the newly injected
artificial cluster. This a simplification, because we 
ignore the magnitudes of the primary and the secondary, but the 
experiments indicated that only a negligibly small fraction of 
sources are merged and they have little effect on the cluster 
identification. This is related to the choice of the prototype 
cluster Westerlund\,2 -- it is already fairly distant at $\sim$4\,kpc;
the impact of the changing viewing geometry would have been much 
more significant if our prototype was closer, for example 
less than a kiloparsec from us, so the movement to the distance 
of the Galactic center would reduce the separations between stars 
by nearly an order of magnitude.

The grid was defined to place the artificial clusters in the 
innermost Galaxy -- the region that is most difficult for cluster 
searches because of both crowding and extinction: we inserted 
clusters in 600 steps within the range 
$-$30$\leq$$l$$\leq$30\,deg and in 12 steps within 
$-$0.5$\leq$$b$$\leq$0.5\,deg, resulting in 7200 grid points.
For each point, we carried out 9 separate simulations introducing
artificial clusters subjected to all possible combinations of 
distances D=6, 7 and 8\,kpc and visual extinctions $A_V$=11, 12 
and 13\,mag, in other words this is a 3$\times$3 grid.
The choice of exact values is tentative, but they were selected
to cover the ranges typical for the other massive clusters in
this region. We limited the distances to span only the near side
of the Milky Way. Higher extinctions have been found for clusters
in dust-rich young star-forming regions
\citep[e.g.,][]{2005A&A...435...95B,2006A&A...455..923B}
but not for super-massive clusters. 

The artificial clusters were added on top of the stellar fields 
at the predefined locations. Unlike in the previous step where 
we merged starts that come too close together as the Westerlund\,2
sources
are moved further out to the adopted new distance, here we ignore 
the merging of the member stars and any field sources that may 
come too close to them, because a check indicated that 
$\sim$2--4\,\% of the suspected members are affected, even in the 
densest regions -- typically less than one star per cluster.

\begin{figure}
\centering
\includegraphics[width=0.9\linewidth]{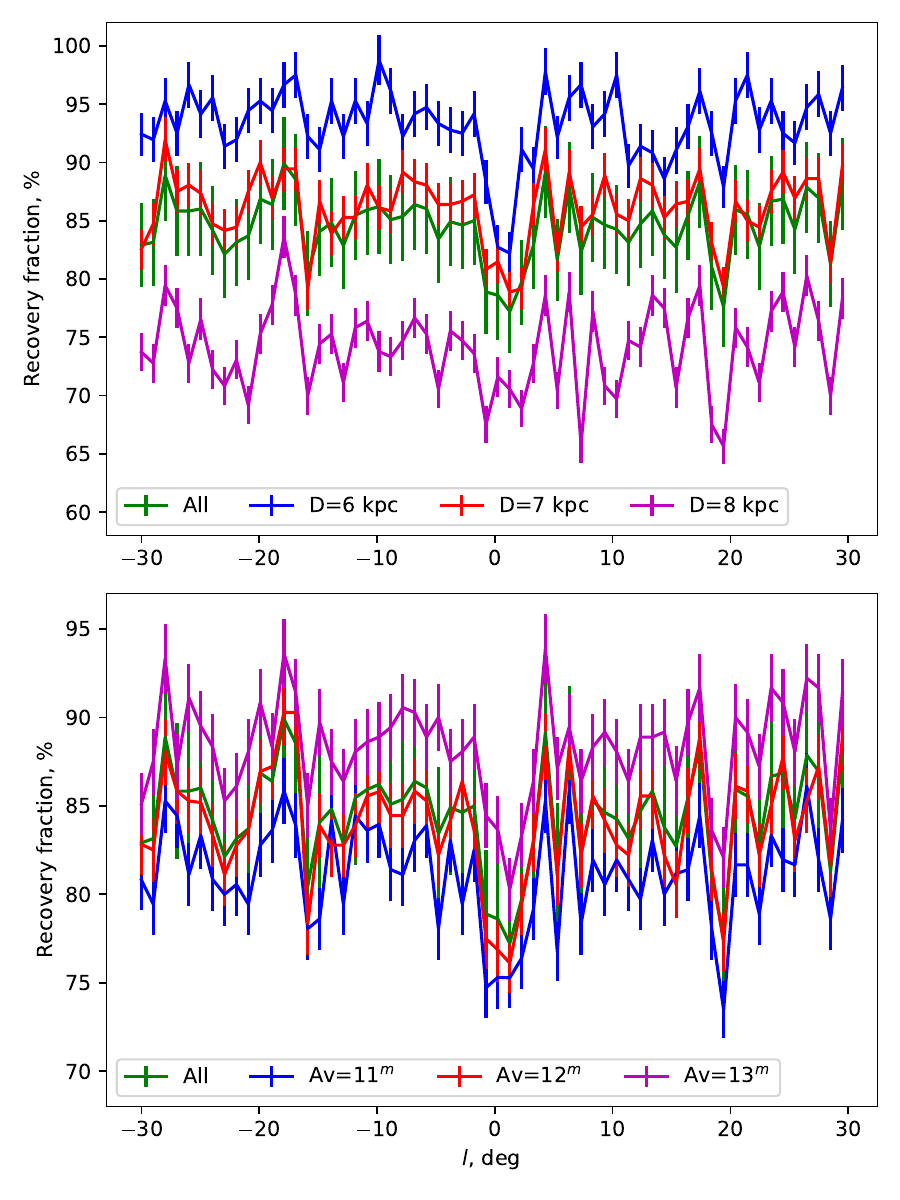}
\includegraphics[width=0.9\linewidth]{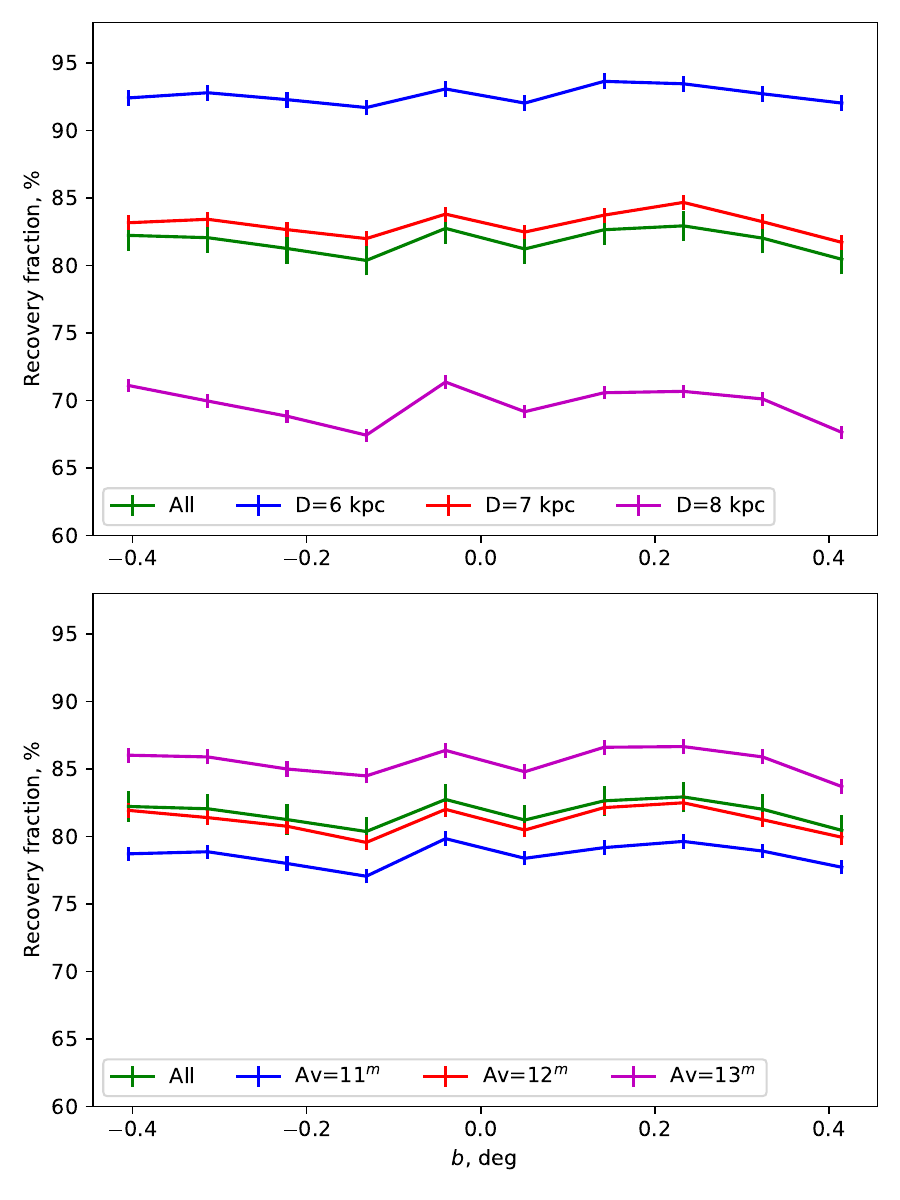}
\caption{Rate of artificial cluster recovery as a function of
galactic coordinates for the entire sample, and for different
distances D and visual extinctions $A_V$ (labeled). For every
value D or $A_V$ we average over the 9-element grid (see 
Sec.\,\ref{sec:gen_clust}) along that value.}
\label{fig:recovery}
\end{figure}

\subsection{Recovery of the artificial clusters}

The same algorithm that was used for the cluster search was 
applied to the catalogs with the artificial clusters and the 
behavior of the recovery rates is shown in 
Fig.\,\ref{fig:recovery}. The fraction of recovered clusters
varies between 70\,\% and 95\,\%. Nearby clusters 
are easier to identify than more distant ones. However, the 
higher extinction seems to help to find clusters -- possibly
because it sets the cluster in color space further apart from 
the contaminating foreground population that shows bluer colors
than the candidate cluster member stars. The foreground dominates
the surface density, but it is removed by the color criterion.
An increase of $A_V$ by 1\,mag roughly raises the recovery
fraction by $\sim$4\,\%. Spatially, the innermost region at
$-$2$\leq$$l$$\leq$2\,deg stands out with a somewhat lower
recovery rate, probably because of the worse crowding near the
Galactic Center. In Galactic latitude, there seems to be no drop
in the recovery rate at the position of the Milky Way plane
within the range of latitude $b$ that is covered by our
simulation.

\section{Discussion and conclusions}\label{sec:conclusions}

We applied a new cluster finding algorithm -- OPTICS -- on the
GLIMPSE survey point source catalog to identify obscured star
clusters located in the inner Milky Way and report nearly 500
new objects; we also recovered about 140 previously identified
ones. Importantly, these are all candidates, because without
spectroscopy, or proper motions, or parallaxes it is impossible
to verify their nature as bound systems. The deep spectroscopic
surveys like Multi-Object Optical and Near-infrared Spectrograph
\citep[MOONS,][]{2020Msngr.180...10C}, 4 metre Multi-Object
Spectroscopic Telescope \citep[4MOST,][]{2019Msngr.175....3D}
and The Wide-field Spectroscopic Telescope
\citep[WST,][]{2024arXiv240305398M}, and the near-IR astrometric
space missions of the next decades will help to address this
question. In particular, the future {\it Roman} Space Telescope
will have wide-field near-IR imaging capabilities
\citep[e.g.,][]{2018arXiv180600554S}, ideal to make a very deep
survey of the Galactic plane at high resolution
\citep[see,][]{2023arXiv230707642P}. Such a survey would not
only discover thousands of new star clusters enabling them to
complete their census, but also to confirm their true nature
using proper motions.

The new OPTICS algorithm can handle hierarchical structure of
the star formation but it proved not to be too important here,
because the severe field contamination in the direction of the
inner Galaxy typically allows us to identify only the most
compact obscured clusters; very few structures larger than
4-5\,arcmin were identified. We speculate that the capability
of the OPTICS algorithm to handle hierarchical structures will
be important for studies of nearby galaxies like the Magellanic
Clouds, M31, and M33. The classification and characterization
of the new candidates remains outside the scope of this work
but the properties of recovered known clusters do hint at the
possibility that most of the new candidates would also be
embedded and maybe a few would be highly obscured open or
globular clusters. The distance estimates of these embedded
clusters are not possible without proper spectroscopic or NIR
astrometric missions and without this knowledge, we cannot
claim that our cluster candidates are bonafide clusters or
not. So, while this new list is a step towards a more complete
observational census of stellar clusters, this goal remains
unattainable as our simulation shows, because the recovery
rate for less massive, less rich, and less luminous clusters
are bound to be lower than our estimates for clusters similar
to Westerlund\,2.

One challenge that remains to be addressed in future work is 
to improve the field contamination removal. The member stars
of the cluster or the cluster candidates are more crowded than
in the surrounding field. Therefore, the luminosity function in
the cluster is shallower than in the field because of the source 
confusion. This may be one of the reasons -- in addition to the 
stochastic or dark cloud-related surface density variations -- 
why do we sometimes have more stars in the sky annulus than in 
the cluster locus, as mentioned in Sec.\,\ref{sec:properties}. 
An inspection of the clusters and field luminosity functions 
indicated that the small number of stars makes it difficult to 
determine the completeness limit and we refrained from 
applying a completeness correction or even just removing 
the faint stars below some limit. This problem is similar to 
the edge detection issue in the tip of the red giant branch 
method to measure extragalactic distances 
\citep{1996ApJ...461..713S}. This method requires a large 
number of stars which prevents its application to globular
clusters. We refrain from applying an uncertain correction
also because the over-subtraction of the field stars reduces 
the recovery rate of our simulation, making our result more
conservative.

Here we addressed the important but often neglected question
of how successful our algorithm is in finding clusters with a
simulation, adding semi-artificial clusters to the GLIMPSE
point source catalog, and running the same search algorithm
trying to recover them. As a first step, the simulation is
limited to the most massive star clusters, with masses around
10$^4$\, M$_\odot$. It is semi-artificial because it is based
on a real cluster -- Westerlund\,2 -- which makes it
model-independent. The simulation is also limited to the
central part of our Galaxy, where the extinction is high and
crowding is severe. The achieved recovery fraction is high --
in the range 70--95\,\%, suggesting that the near side of the
Milky Way may harbor $\sim$1-3 additional supermassive star
clusters. In other words, no large population of hidden s
upermassive clusters resides inside the Milky Way. The analysis
of the simulated clusters indicates that the closer ones are
easier to identify than their more distant counterparts, but
the higher extinction often helps to identify clusters, because
it increased the color contrast between them and the
contamination field population.

Our simulation is the first detailed one, after the initial
attempts by \citet{2005ApJ...635..560M}, \citet{2010IAUS..266..203I}
and \citet{2010HiA....15..794H}. It needs to be extended to
include lower mass and older clusters that have a larger impact
on the stellar population in our Galaxy because of their higher
numbers than the Westerlund\,2-like clusters. An expansion
towards near-IR sky surveys that typically have better angular
resolution than the mid-IR surveys is another promising expansion
avenue, for example, the VVV/X \citep[][]{2024A&A...689A.148S, 2010NewA...15..433M}.

Last but not least, we underline again that the objects that
the OPTICS algorithm identified are candidates and further
observations are needed to confirm their cluster nature.

\begin{acknowledgements}
This research has made use of the SIMBAD database, operated at
CDS, Strasbourg, France. This research has made use of the
VizieR catalog access tool, CDS, Strasbourg, France
\citep{10.26093/cds/vizier}. The original description of the
VizieR service was published in \citet{vizier2000}. This
research has made use of the NASA/IPAC Infrared Science
Archive, which is funded by the National Aeronautics and Space
Administration and operated by the California Institute of
Technology. The authors are grateful to LMU and ESO for the
usage of computational facilities. The research of T.P. was
supported by the Excellence Cluster ORIGINS, which is funded
by the Deutsche Forschungsgemeinschaft (DFG, German Research
Foundation) under Germany's Excellence
Strategy-EXC-2094-390783311.
D.M. gratefully acknowledges support from the Center for
Astrophysics and Associated Technologies CATA by the ANID
BASAL projects ACE210002 and FB210003, and by Fondecyt Regular
No. 1220724.
We thank the anonymous referee for the
helpful comments that helped to improve the manuscript greatly. 
\end{acknowledgements}

\bibliographystyle{aa}
\bibliography{bib}

\begin{thebibliography}{73}
\expandafter\ifx\csname natexlab\endcsname\relax\def\natexlab#1{#1}\fi

\bibitem[{Ankerst {et~al.}(1999)Ankerst, Breunig, Kriegel, \&
  Sander}]{10.1145/304181.304187}
Ankerst, M., Breunig, M.~M., Kriegel, H.-P., \& Sander, J. 1999, SIGMOD Rec.,
  28, 49–60

\bibitem[{{Asa'd} {et~al.}(2023){Asa'd}, {Ivanov}, {Negueruela}, {John},
  {Gonneau}, \& {Rejkuba}}]{2023AJ....165..212A}
{Asa'd}, R., {Ivanov}, V.~D., {Negueruela}, I., {et~al.} 2023, \aj, 165, 212

\bibitem[{{Astropy Collaboration} {et~al.}(2022){Astropy Collaboration},
  {Price-Whelan}, {Lim}, {Earl}, {Starkman}, {Bradley}, {Shupe}, {Patil},
  {Corrales}, {Brasseur}, {N{\"o}the}, {Donath}, {Tollerud}, {Morris},
  {Ginsburg}, {Vaher}, {Weaver}, {Tocknell}, {Jamieson}, {van Kerkwijk},
  {Robitaille}, {Merry}, {Bachetti}, {G{\"u}nther}, {Aldcroft},
  {Alvarado-Montes}, {Archibald}, {B{\'o}di}, {Bapat}, {Barentsen},
  {Baz{\'a}n}, {Biswas}, {Boquien}, {Burke}, {Cara}, {Cara}, {Conroy},
  {Conseil}, {Craig}, {Cross}, {Cruz}, {D'Eugenio}, {Dencheva}, {Devillepoix},
  {Dietrich}, {Eigenbrot}, {Erben}, {Ferreira}, {Foreman-Mackey}, {Fox},
  {Freij}, {Garg}, {Geda}, {Glattly}, {Gondhalekar}, {Gordon}, {Grant},
  {Greenfield}, {Groener}, {Guest}, {Gurovich}, {Handberg}, {Hart},
  {Hatfield-Dodds}, {Homeier}, {Hosseinzadeh}, {Jenness}, {Jones}, {Joseph},
  {Kalmbach}, {Karamehmetoglu}, {Ka{\l}uszy{\'n}ski}, {Kelley}, {Kern},
  {Kerzendorf}, {Koch}, {Kulumani}, {Lee}, {Ly}, {Ma}, {MacBride}, {Maljaars},
  {Muna}, {Murphy}, {Norman}, {O'Steen}, {Oman}, {Pacifici}, {Pascual},
  {Pascual-Granado}, {Patil}, {Perren}, {Pickering}, {Rastogi}, {Roulston},
  {Ryan}, {Rykoff}, {Sabater}, {Sakurikar}, {Salgado}, {Sanghi}, {Saunders},
  {Savchenko}, {Schwardt}, {Seifert-Eckert}, {Shih}, {Jain}, {Shukla}, {Sick},
  {Simpson}, {Singanamalla}, {Singer}, {Singhal}, {Sinha}, {Sip{\H{o}}cz},
  {Spitler}, {Stansby}, {Streicher}, {{\v{S}}umak}, {Swinbank}, {Taranu},
  {Tewary}, {Tremblay}, {de Val-Borro}, {Van Kooten}, {Vasovi{\'c}}, {Verma},
  {de Miranda Cardoso}, {Williams}, {Wilson}, {Winkel}, {Wood-Vasey}, {Xue},
  {Yoachim}, {Zhang}, {Zonca}, \& {Astropy Project
  Contributors}}]{2022ApJ...935..167A}
{Astropy Collaboration}, {Price-Whelan}, A.~M., {Lim}, P.~L., {et~al.} 2022,
  \apj, 935, 167

\bibitem[{{Astropy Collaboration} {et~al.}(2018){Astropy Collaboration},
  {Price-Whelan}, {Sip{\H{o}}cz}, {G{\"u}nther}, {Lim}, {Crawford}, {Conseil},
  {Shupe}, {Craig}, {Dencheva}, {Ginsburg}, {VanderPlas}, {Bradley},
  {P{\'e}rez-Su{\'a}rez}, {de Val-Borro}, {Aldcroft}, {Cruz}, {Robitaille},
  {Tollerud}, {Ardelean}, {Babej}, {Bach}, {Bachetti}, {Bakanov}, {Bamford},
  {Barentsen}, {Barmby}, {Baumbach}, {Berry}, {Biscani}, {Boquien}, {Bostroem},
  {Bouma}, {Brammer}, {Bray}, {Breytenbach}, {Buddelmeijer}, {Burke},
  {Calderone}, {Cano Rodr{\'\i}guez}, {Cara}, {Cardoso}, {Cheedella}, {Copin},
  {Corrales}, {Crichton}, {D'Avella}, {Deil}, {Depagne}, {Dietrich}, {Donath},
  {Droettboom}, {Earl}, {Erben}, {Fabbro}, {Ferreira}, {Finethy}, {Fox},
  {Garrison}, {Gibbons}, {Goldstein}, {Gommers}, {Greco}, {Greenfield},
  {Groener}, {Grollier}, {Hagen}, {Hirst}, {Homeier}, {Horton}, {Hosseinzadeh},
  {Hu}, {Hunkeler}, {Ivezi{\'c}}, {Jain}, {Jenness}, {Kanarek}, {Kendrew},
  {Kern}, {Kerzendorf}, {Khvalko}, {King}, {Kirkby}, {Kulkarni}, {Kumar},
  {Lee}, {Lenz}, {Littlefair}, {Ma}, {Macleod}, {Mastropietro}, {McCully},
  {Montagnac}, {Morris}, {Mueller}, {Mumford}, {Muna}, {Murphy}, {Nelson},
  {Nguyen}, {Ninan}, {N{\"o}the}, {Ogaz}, {Oh}, {Parejko}, {Parley}, {Pascual},
  {Patil}, {Patil}, {Plunkett}, {Prochaska}, {Rastogi}, {Reddy Janga},
  {Sabater}, {Sakurikar}, {Seifert}, {Sherbert}, {Sherwood-Taylor}, {Shih},
  {Sick}, {Silbiger}, {Singanamalla}, {Singer}, {Sladen}, {Sooley},
  {Sornarajah}, {Streicher}, {Teuben}, {Thomas}, {Tremblay}, {Turner},
  {Terr{\'o}n}, {van Kerkwijk}, {de la Vega}, {Watkins}, {Weaver}, {Whitmore},
  {Woillez}, {Zabalza}, \& {Astropy Contributors}}]{2018AJ....156..123A}
{Astropy Collaboration}, {Price-Whelan}, A.~M., {Sip{\H{o}}cz}, B.~M., {et~al.}
  2018, \aj, 156, 123

\bibitem[{{Barb{\'a}} {et~al.}(2015){Barb{\'a}}, {Roman-Lopes}, {Nilo
  Castell{\'o}n}, {Firpo}, {Minniti}, {Lucas}, {Emerson}, {Hempel}, {Soto}, \&
  {Saito}}]{2015A&A...581A.120B}
{Barb{\'a}}, R.~H., {Roman-Lopes}, A., {Nilo Castell{\'o}n}, J.~L., {et~al.}
  2015, \aap, 581, A120

\bibitem[{{Becklin} \& {Neugebauer}(1968)}]{1968ApJ...151..145B}
{Becklin}, E.~E. \& {Neugebauer}, G. 1968, \apj, 151, 145

\bibitem[{{Benjamin} {et~al.}(2003){Benjamin}, {Churchwell}, {Babler}, {Bania},
  {Clemens}, {Cohen}, {Dickey}, {Indebetouw}, {Jackson}, {Kobulnicky},
  {Lazarian}, {Marston}, {Mathis}, {Meade}, {Seager}, {Stolovy}, {Watson},
  {Whitney}, {Wolff}, \& {Wolfire}}]{2003PASP..115..953B}
{Benjamin}, R.~A., {Churchwell}, E., {Babler}, B.~L., {et~al.} 2003, \pasp,
  115, 953

\bibitem[{{Bica} {et~al.}(2003){Bica}, {Dutra}, {Soares}, \&
  {Barbuy}}]{2003A&A...404..223B}
{Bica}, E., {Dutra}, C.~M., {Soares}, J., \& {Barbuy}, B. 2003, \aap, 404, 223

\bibitem[{{Borissova} {et~al.}(2011){Borissova}, {Bonatto}, {Kurtev}, {Clarke},
  {Pe{\~n}aloza}, {Sale}, {Minniti}, {Alonso-Garc{\'\i}a}, {Artigau},
  {Barb{\'a}}, {Bica}, {Baume}, {Catelan}, {Chen{\`e}}, {Dias}, {Folkes},
  {Froebrich}, {Geisler}, {de Grijs}, {Hanson}, {Hempel}, {Ivanov}, {Kumar},
  {Lucas}, {Mauro}, {Moni Bidin}, {Rejkuba}, {Saito}, {Tamura}, \&
  {Toledo}}]{2011A&A...532A.131B}
{Borissova}, J., {Bonatto}, C., {Kurtev}, R., {et~al.} 2011, \aap, 532, A131

\bibitem[{{Borissova} {et~al.}(2018){Borissova}, {Ivanov}, {Lucas}, {Kurtev},
  {Alonso-Garcia}, {Ram{\'\i}rez Alegr{\'\i}a}, {Minniti}, {Froebrich},
  {Hempel}, {Medina}, {Chen{\'e}}, \& {Kuhn}}]{2018MNRAS.481.3902B}
{Borissova}, J., {Ivanov}, V.~D., {Lucas}, P.~W., {et~al.} 2018, \mnras, 481,
  3902

\bibitem[{{Borissova} {et~al.}(2006){Borissova}, {Ivanov}, {Minniti}, \&
  {Geisler}}]{2006A&A...455..923B}
{Borissova}, J., {Ivanov}, V.~D., {Minniti}, D., \& {Geisler}, D. 2006, \aap,
  455, 923

\bibitem[{{Borissova} {et~al.}(2005){Borissova}, {Ivanov}, {Minniti},
  {Geisler}, \& {Stephens}}]{2005A&A...435...95B}
{Borissova}, J., {Ivanov}, V.~D., {Minniti}, D., {Geisler}, D., \& {Stephens},
  A.~W. 2005, \aap, 435, 95

\bibitem[{{Camargo} {et~al.}(2016){Camargo}, {Bica}, \&
  {Bonatto}}]{2016MNRAS.455.3126C}
{Camargo}, D., {Bica}, E., \& {Bonatto}, C. 2016, \mnras, 455, 3126

\bibitem[{{Cirasuolo} {et~al.}(2020){Cirasuolo}, {Fairley}, {Rees}, {Gonzalez},
  {Taylor}, {Maiolino}, {Afonso}, {Evans}, {Flores}, {Lilly}, {Oliva},
  {Paltani}, {Vanzi}, {Abreu}, {Accardo}, {Adams}, {{\'A}lvarez M{\'e}ndez},
  {Amans}, {Amarantidis}, {Atek}, {Atkinson}, {Banerji}, {Barrett},
  {Barrientos}, {Bauer}, {Beard}, {B{\'e}chet}, {Belfiore}, {Bellazzini},
  {Benoist}, {Best}, {Biazzo}, {Black}, {Boettger}, {Bonifacio}, {Bowler},
  {Bragaglia}, {Brierley}, {Brinchmann}, {Brinkmann}, {Buat}, {Buitrago},
  {Burgarella}, {Burningham}, {Buscher}, {Cabral}, {Caffau}, {Cardoso},
  {Carnall}, {Carollo}, {Castillo}, {Castignani}, {Catelan}, {Cicone},
  {Cimatti}, {Cioni}, {Clementini}, {Cochrane}, {Coelho}, {Colling}, {Contini},
  {Contreras}, {Conzelmann}, {Cresci}, {Cropper}, {Cucciati}, {Cullen},
  {Cumani}, {Curti}, {Da Silva}, {Daddi}, {Dalessandro}, {Dalessio}, {Dauvin},
  {Davidson}, {de Laverny}, {Delplancke-Str{\"o}bele}, {De Lucia}, {Del
  Vecchio}, {Dessauges-Zavadsky}, {Di Matteo}, {Dole}, {Drass}, {Dunlop},
  {D{\"u}nner}, {Eales}, {Ellis}, {Enriques}, {Fasola}, {Ferguson}, {Ferruzzi},
  {Fisher}, {Flores}, {Fontana}, {Forchi}, {Francois}, {Franzetti}, {Gargiulo},
  {Garilli}, {Gaudemard}, {Gieles}, {Gilmore}, {Ginolfi}, {Gomes}, {Guinouard},
  {Gutierrez}, {Haigron}, {Hammer}, {Hammersley}, {Haniff}, {Harrison},
  {Haywood}, {Hill}, {Hubin}, {Humphrey}, {Ibata}, {Infante}, {Ives}, {Ivison},
  {Iwert}, {Jablonka}, {Jakob}, {Jarvis}, {King}, {Kneib}, {Laporte},
  {Lawrence}, {Lee}, {Li Causi}, {Lorenzoni}, {Lucatello}, {Luco}, {Macleod},
  {Magliocchetti}, {Magrini}, {Mainieri}, {Maire}, {Mannucci}, {Martin},
  {Matute}, {Maurogordato}, {McGee}, {Mcleod}, {McLure}, {McMahon}, {Melse},
  {Messias}, {Mucciarelli}, {Nisini}, {Nix}, {Norberg}, {Oesch}, {Oliveira},
  {Origlia}, {Padilla}, {Palsa}, {Pancino}, {Papaderos}, {Pappalardo}, {Parry},
  {Pasquini}, {Peacock}, {Pedichini}, {Pello}, {Peng}, {Pentericci}, {Pfuhl},
  {Piazzesi}, {Popovic}, {Pozzetti}, {Puech}, {Puzia}, {Raichoor}, {Randich},
  {Recio-Blanco}, {Reis}, {Reix}, {Renzini}, {Rodrigues}, {Rojas},
  {Rojas-Arriagada}, {Rota}, {Royer}, {Sacco}, {Sanchez-Janssen}, {Sanna},
  {Santos}, {Sarzi}, {Schaerer}, {Schiavon}, {Schnell}, {Schultheis},
  {Scodeggio}, {Serjeant}, {Shen}, {Simmonds}, {Smoker}, {Sobral}, {Sordet},
  {Sp{\'e}rone}, {Strachan}, {Sun}, {Swinbank}, {Tait}, {Tereno}, {Tojeiro},
  {Torres}, {Tosi}, {Tozzi}, {Tresiter}, {Valenti}, {Valenzuela Navarro},
  {Vanzella}, {Vergani}, {Verhamme}, {Vernet}, {Vignali}, {Vinther}, {Von
  Dran}, {Waring}, {Watson}, {Wild}, {Willesme}, {Woodward}, {Wuyts}, {Yang},
  {Zamorani}, {Zoccali}, {Bluck}, \& {Trussler}}]{2020Msngr.180...10C}
{Cirasuolo}, M., {Fairley}, A., {Rees}, P., {et~al.} 2020, The Messenger, 180,
  10

\bibitem[{{Contreras} {et~al.}(2013){Contreras}, {Schuller}, {Urquhart},
  {Csengeri}, {Wyrowski}, {Beuther}, {Bontemps}, {Bronfman}, {Henning},
  {Menten}, {Schilke}, {Walmsley}, {Wienen}, {Tackenberg}, \&
  {Linz}}]{2013A&A...549A..45C}
{Contreras}, Y., {Schuller}, F., {Urquhart}, J.~S., {et~al.} 2013, \aap, 549,
  A45

\bibitem[{{Davies} {et~al.}(2007){Davies}, {Figer}, {Kudritzki}, {MacKenty},
  {Najarro}, \& {Herrero}}]{2007ApJ...671..781D}
{Davies}, B., {Figer}, D.~F., {Kudritzki}, R.-P., {et~al.} 2007, \apj, 671, 781

\bibitem[{{de Jong} {et~al.}(2019){de Jong}, {Agertz}, {Berbel}, {Aird},
  {Alexander}, {Amarsi}, {Anders}, {Andrae}, {Ansarinejad}, {Ansorge},
  {Antilogus}, {Anwand-Heerwart}, {Arentsen}, {Arnadottir}, {Asplund}, {Auger},
  {Azais}, {Baade}, {Baker}, {Baker}, {Balbinot}, {Baldry}, {Banerji},
  {Barden}, {Barklem}, {Barth{\'e}l{\'e}my-Mazot}, {Battistini}, {Bauer},
  {Bell}, {Bellido-Tirado}, {Bellstedt}, {Belokurov}, {Bensby}, {Bergemann},
  {Bestenlehner}, {Bielby}, {Bilicki}, {Blake}, {Bland-Hawthorn}, {Boeche},
  {Boland}, {Boller}, {Bongard}, {Bongiorno}, {Bonifacio}, {Boudon}, {Brooks},
  {Brown}, {Brown}, {Br{\"u}ggen}, {Brynnel}, {Brzeski}, {Buchert},
  {Buschkamp}, {Caffau}, {Caillier}, {Carrick}, {Casagrande}, {Case}, {Casey},
  {Cesarini}, {Cescutti}, {Chapuis}, {Chiappini}, {Childress}, {Christlieb},
  {Church}, {Cioni}, {Cluver}, {Colless}, {Collett}, {Comparat}, {Cooper},
  {Couch}, {Courbin}, {Croom}, {Croton}, {Daguis{\'e}}, {Dalton}, {Davies},
  {Davis}, {de Laverny}, {Deason}, {Dionies}, {Disseau}, {Doel}, {D{\"o}scher},
  {Driver}, {Dwelly}, {Eckert}, {Edge}, {Edvardsson}, {Youssoufi}, {Elhaddad},
  {Enke}, {Erfanianfar}, {Farrell}, {Fechner}, {Feiz}, {Feltzing}, {Ferreras},
  {Feuerstein}, {Feuillet}, {Finoguenov}, {Ford}, {Fotopoulou}, {Fouesneau},
  {Frenk}, {Frey}, {Gaessler}, {Geier}, {Gentile Fusillo}, {Gerhard},
  {Giannantonio}, {Giannone}, {Gibson}, {Gillingham},
  {Gonz{\'a}lez-Fern{\'a}ndez}, {Gonzalez-Solares}, {Gottloeber}, {Gould},
  {Grebel}, {Gueguen}, {Guiglion}, {Haehnelt}, {Hahn}, {Hansen}, {Hartman},
  {Hauptner}, {Hawkins}, {Haynes}, {Haynes}, {Heiter}, {Helmi}, {Aguayo},
  {Hewett}, {Hinton}, {Hobbs}, {Hoenig}, {Hofman}, {Hook}, {Hopgood},
  {Hopkins}, {Hourihane}, {Howes}, {Howlett}, {Huet}, {Irwin}, {Iwert},
  {Jablonka}, {Jahn}, {Jahnke}, {Jarno}, {Jin}, {Jofre}, {Johl}, {Jones},
  {J{\"o}nsson}, {Jordan}, {Karovicova}, {Khalatyan}, {Kelz}, {Kennicutt},
  {King}, {Kitaura}, {Klar}, {Klauser}, {Kneib}, {Koch}, {Koposov},
  {Kordopatis}, {Korn}, {Kosmalski}, {Kotak}, {Kovalev}, {Kreckel}, {Kripak},
  {Krumpe}, {Kuijken}, {Kunder}, {Kushniruk}, {Lam}, {Lamer}, {Laurent},
  {Lawrence}, {Lehmitz}, {Lemasle}, {Lewis}, {Li}, {Lidman}, {Lind}, {Liske},
  {Lizon}, {Loveday}, {Ludwig}, {McDermid}, {Maguire}, {Mainieri}, {Mali},
  {Mandel}, {Mandel}, {Mannering}, {Martell}, {Martinez Delgado}, {Matijevic},
  {McGregor}, {McMahon}, {McMillan}, {Mena}, {Merloni}, {Meyer}, {Michel},
  {Micheva}, {Migniau}, {Minchev}, {Monari}, {Muller}, {Murphy},
  {Muthukrishna}, {Nandra}, {Navarro}, {Ness}, {Nichani}, {Nichol}, {Nicklas},
  {Niederhofer}, {Norberg}, {Obreschkow}, {Oliver}, {Owers}, {Pai},
  {Pankratow}, {Parkinson}, {Paschke}, {Paterson}, {Pecontal}, {Parry},
  {Phillips}, {Pillepich}, {Pinard}, {Pirard}, {Piskunov}, {Plank},
  {Pl{\"u}schke}, {Pons}, {Popesso}, {Power}, {Pragt}, {Pramskiy}, {Pryer},
  {Quattri}, {Queiroz}, {Quirrenbach}, {Rahurkar}, {Raichoor}, {Ramstedt},
  {Rau}, {Recio-Blanco}, {Reiss}, {Renaud}, {Revaz}, {Rhode}, {Richard},
  {Richter}, {Rix}, {Robotham}, {Roelfsema}, {Romaniello}, {Rosario},
  {Rothmaier}, {Roukema}, {Ruchti}, {Rupprecht}, {Rybizki}, {Ryde}, {Saar},
  {Sadler}, {Sahl{\'e}n}, {Salvato}, {Sassolas}, {Saunders}, {Saviauk},
  {Sbordone}, {Schmidt}, {Schnurr}, {Scholz}, {Schwope}, {Seifert}, {Shanks},
  {Sheinis}, {Sivov}, {Sk{\'u}lad{\'o}ttir}, {Smartt}, {Smedley}, {Smith},
  {Smith}, {Sorce}, {Spitler}, {Starkenburg}, {Steinmetz}, {Stilz}, {Storm},
  {Sullivan}, {Sutherland}, {Swann}, {Tamone}, {Taylor}, {Teillon}, {Tempel},
  {ter Horst}, {Thi}, {Tolstoy}, {Trager}, {Traven}, {Tremblay}, {Tresse},
  {Valentini}, {van de Weygaert}, {van den Ancker}, {Veljanoski}, {Venkatesan},
  {Wagner}, {Wagner}, {Walcher}, {Waller}, {Walton}, {Wang}, {Winkler},
  {Wisotzki}, {Worley}, {Worseck}, {Xiang}, {Xu}, {Yong}, {Zhao}, {Zheng},
  {Zscheyge}, \& {Zucker}}]{2019Msngr.175....3D}
{de Jong}, R.~S., {Agertz}, O., {Berbel}, A.~A., {et~al.} 2019, The Messenger,
  175, 3

\bibitem[{{Di Francesco} {et~al.}(2008){Di Francesco}, {Johnstone}, {Kirk},
  {MacKenzie}, \& {Ledwosinska}}]{2008ApJS..175..277D}
{Di Francesco}, J., {Johnstone}, D., {Kirk}, H., {MacKenzie}, T., \&
  {Ledwosinska}, E. 2008, \apjs, 175, 277

\bibitem[Saito et al.(2024)]{2024A&A...689A.148S} Saito, R.~K., Hempel, M., Alonso-Garc{\'\i}a, J., et al.\ 2024, \aap, 689, A148. doi:10.1051/0004-6361/202450584

\bibitem[{{Dutra} \& {Bica}(2001)}]{2001A&A...376..434D}
{Dutra}, C.~M. \& {Bica}, E. 2001, \aap, 376, 434

\bibitem[{Ester {et~al.}(1996)Ester, Kriegel, Sander, \&
  Xu}]{10.5555/3001460.3001507}
Ester, M., Kriegel, H.-P., Sander, J., \& Xu, X. 1996, in Proceedings of the
  Second International Conference on Knowledge Discovery and Data Mining,
  KDD'96 (AAAI Press), 226–231

\bibitem[{{Evans} {et~al.}(2010){Evans}, {Primini}, {Glotfelty}, {Anderson},
  {Bonaventura}, {Chen}, {Davis}, {Doe}, {Evans}, {Fabbiano}, {Galle}, {Gibbs},
  {Grier}, {Hain}, {Hall}, {Harbo}, {He}, {Houck}, {Karovska}, {Kashyap},
  {Lauer}, {McCollough}, {McDowell}, {Miller}, {Mitschang}, {Morgan},
  {Mossman}, {Nichols}, {Nowak}, {Plummer}, {Refsdal}, {Rots}, {Siemiginowska},
  {Sundheim}, {Tibbetts}, {Van Stone}, {Winkelman}, \&
  {Zografou}}]{2010ApJS..189...37E}
{Evans}, I.~N., {Primini}, F.~A., {Glotfelty}, K.~J., {et~al.} 2010, \apjs,
  189, 37

\bibitem[{{Feigelson} {et~al.}(2007){Feigelson}, {Townsley}, {G{\"u}del}, \&
  {Stassun}}]{2007prpl.conf..313F}
{Feigelson}, E., {Townsley}, L., {G{\"u}del}, M., \& {Stassun}, K. 2007, in
  Protostars and Planets V, ed. B.~{Reipurth}, D.~{Jewitt}, \& K.~{Keil}, 313

\bibitem[{{Foster} {et~al.}(2011){Foster}, {Jackson}, {Barnes}, {Barris},
  {Brooks}, {Cunningham}, {Finn}, {Fuller}, {Longmore}, {Mascoop}, {Peretto},
  {Rathborne}, {Sanhueza}, {Schuller}, \& {Wyrowski}}]{2011ApJS..197...25F}
{Foster}, J.~B., {Jackson}, J.~M., {Barnes}, P.~J., {et~al.} 2011, \apjs, 197,
  25

\bibitem[{{Hanson} {et~al.}(2010){Hanson}, {Popescu}, {Larsen}, \&
  {Ivanov}}]{2010HiA....15..794H}
{Hanson}, M.~M., {Popescu}, B., {Larsen}, S.~S., \& {Ivanov}, V.~D. 2010,
  Highlights of Astronomy, 15, 794

\bibitem[{{Hoq} {et~al.}(2013){Hoq}, {Jackson}, {Foster}, {Sanhueza},
  {Guzm{\'a}n}, {Whitaker}, {Claysmith}, {Rathborne}, {Vasyunina}, \&
  {Vasyunin}}]{2013ApJ...777..157H}
{Hoq}, S., {Jackson}, J.~M., {Foster}, J.~B., {et~al.} 2013, \apj, 777, 157

\bibitem[{Hunter(2007)}]{Hunter:2007}
Hunter, J.~D. 2007, Computing in Science \& Engineering, 9, 90

\bibitem[{{Hurt} {et~al.}(2000){Hurt}, {Jarrett}, {Kirkpatrick}, {Cutri},
  {Schneider}, {Skrutskie}, \& {van Driel}}]{2000AJ....120.1876H}
{Hurt}, R.~L., {Jarrett}, T.~H., {Kirkpatrick}, J.~D., {et~al.} 2000, \aj, 120,
  1876

\bibitem[{{Ivanov} {et~al.}(2002){Ivanov}, {Borissova}, {Pessev}, {Ivanov}, \&
  {Kurtev}}]{2002A&A...394L...1I}
{Ivanov}, V.~D., {Borissova}, J., {Pessev}, P., {Ivanov}, G.~R., \& {Kurtev},
  R. 2002, \aap, 394, L1

\bibitem[{{Ivanov} {et~al.}(2005){Ivanov}, {Kurtev}, \&
  {Borissova}}]{2005A&A...442..195I}
{Ivanov}, V.~D., {Kurtev}, R., \& {Borissova}, J. 2005, \aap, 442, 195

\bibitem[{{Ivanov} {et~al.}(2010){Ivanov}, {Messineo}, {Zhu}, {Figer},
  {Borissova}, {Kurtev}, \& {Ivanov}}]{2010IAUS..266..203I}
{Ivanov}, V.~D., {Messineo}, M., {Zhu}, Q., {et~al.} 2010, in Star Clusters:
  Basic Galactic Building Blocks Throughout Time and Space, ed. R.~{de Grijs}
  \& J.~R.~D. {L{\'e}pine}, Vol. 266, 203--210

\bibitem[{{Jackson} {et~al.}(2013){Jackson}, {Rathborne}, {Foster}, {Whitaker},
  {Sanhueza}, {Claysmith}, {Mascoop}, {Wienen}, {Breen}, {Herpin},
  {Duarte-Cabral}, {Csengeri}, {Longmore}, {Contreras}, {Indermuehle},
  {Barnes}, {Walsh}, {Cunningham}, {Brooks}, {Britton}, {Voronkov}, {Urquhart},
  {Alves}, {Jordan}, {Hill}, {Hoq}, {Finn}, {Bains}, {Bontemps}, {Bronfman},
  {Caswell}, {Deharveng}, {Ellingsen}, {Fuller}, {Garay}, {Green}, {Hindson},
  {Jones}, {Lenfestey}, {Lo}, {Lowe}, {Mardones}, {Menten}, {Minier}, {Morgan},
  {Motte}, {Muller}, {Peretto}, {Purcell}, {Schilke}, {Bontemps}, {Schuller},
  {Titmarsh}, {Wyrowski}, \& {Zavagno}}]{2013PASA...30...57J}
{Jackson}, J.~M., {Rathborne}, J.~M., {Foster}, J.~B., {et~al.} 2013, \pasa,
  30, e057

\bibitem[{{Kaltcheva} \& {Georgiev}(1993)}]{1993MNRAS.261..847K}
{Kaltcheva}, N.~T. \& {Georgiev}, L.~N. 1993, \mnras, 261, 847

\bibitem[{{Koposov} {et~al.}(2017){Koposov}, {Belokurov}, \&
  {Torrealba}}]{2017MNRAS.470.2702K}
{Koposov}, S.~E., {Belokurov}, V., \& {Torrealba}, G. 2017, \mnras, 470, 2702

\bibitem[{{Kurtev} {et~al.}(2008){Kurtev}, {Ivanov}, {Borissova}, \&
  {Ortolani}}]{2008A&A...489..583K}
{Kurtev}, R., {Ivanov}, V.~D., {Borissova}, J., \& {Ortolani}, S. 2008, \aap,
  489, 583

\bibitem[{{Lada} \& {Lada}(2003)}]{2003ARA&A..41...57L}
{Lada}, C.~J. \& {Lada}, E.~A. 2003, \araa, 41, 57

\bibitem[{{Lin} {et~al.}(2019){Lin}, {Csengeri}, {Wyrowski}, {Urquhart},
  {Schuller}, {Weiss}, \& {Menten}}]{2019A&A...631A..72L}
{Lin}, Y., {Csengeri}, T., {Wyrowski}, F., {et~al.} 2019, \aap, 631, A72

\bibitem[{{Mainieri} {et~al.}(2024){Mainieri}, {Anderson}, {Brinchmann},
  {Cimatti}, {Ellis}, {Hill}, {Kneib}, {McLeod}, {Opitom}, {Roth},
  {Sanchez-Saez}, {Smiljanic}, {Tolstoy}, {Bacon}, {Randich}, {Adamo},
  {Annibali}, {Arevalo}, {Audard}, {Barsanti}, {Battaglia}, {Bayo Aran},
  {Belfiore}, {Bellazzini}, {Bellini}, {Beltran}, {Berni}, {Bianchi}, {Biazzo},
  {Bisero}, {Bisogni}, {Bland-Hawthorn}, {Blondin}, {Bodensteiner}, {Boffin},
  {Bonito}, {Bono}, {Bouche}, {Bowman}, {Braga}, {Bragaglia}, {Branchesi},
  {Brucalassi}, {Bryant}, {Bryson}, {Busa}, {Camera}, {Carbone}, {Casali},
  {Casali}, {Casasola}, {Castro}, {Catelan}, {Cavallo}, {Chiappini}, {Cioni},
  {Colless}, {Colzi}, {Contarini}, {Couch}, {D'Ammando}, {d'Assignies D.},
  {D'Orazi}, {da Silva}, {Dainotti}, {Damiani}, {Danielski}, {De Cia}, {de
  Jong}, {Dhawan}, {Dierickx}, {Driver}, {Dupletsa}, {Escoffier}, {Escorza},
  {Fabrizio}, {Fiorentino}, {Fontana}, {Fontani}, {Forero Sanchez}, {Franois},
  {Galindo-Guil}, {Gallazzi}, {Galli}, {Garcia}, {Garcia-Rojas}, {Garilli},
  {Grand}, {Guarcello}, {Hazra}, {Helmi}, {Herrero}, {Iglesias}, {Ilic},
  {Irsic}, {Ivanov}, {Izzo}, {Jablonka}, {Joachimi}, {Kakkad}, {Kamann},
  {Koposov}, {Kordopatis}, {Kovacevic}, {Kraljic}, {Kuncarayakti}, {Kwon}, {La
  Forgia}, {Lahav}, {Laigle}, {Lazzarin}, {Leaman}, {Leclercq}, {Lee}, {Lee},
  {Lehnert}, {Lira}, {Loffredo}, {Lucatello}, {Magrini}, {Maguire}, {Mahler},
  {Zahra Majidi}, {Malavasi}, {Mannucci}, {Marconi}, {Martin}, {Marulli},
  {Massari}, {Matsuno}, {Mattheee}, {McGee}, {Merc}, {Merle}, {Miglio},
  {Migliorini}, {Minchev}, {Minniti}, {Miret-Roig}, {Monreal Ibero}, {Montano},
  {Montet}, {Moresco}, {Moretti}, {Moscardini}, {Moya}, {Mueller},
  {Nanayakkara}, {Nicholl}, {Nordlander}, {Onori}, {Padovani}, {Pala}, {Panda},
  {Pandey-Pommier}, {Pasquini}, {Pawlak}, {Pessi}, {Pisani}, {Popovic},
  {Prisinzano}, {Raddi}, {Rainer}, {Rebassa-Mansergas}, {Richard}, {Rigault},
  {Rocher}, {Romano}, {Rosati}, {Sacco}, {Sanchez-Janssen}, {Sander},
  {Sanders}, {Sargent}, {Sarpa}, {Schimd}, {Schipani}, {Sefusatti}, {Smith},
  {Spina}, {Steinmetz}, {Tacchella}, {Tautvaisiene}, {Theissen}, {Thomas},
  {Ting}, {Travouillon}, {Tresse}, {Trivedi}, {Tsantaki}, {Tsedrik}, {Urrutia},
  {Valenti}, {Van der Swaelmen}, {Van Eck}, {Verdiani}, {Verdier}, {Vergani},
  {Verhamme}, {Vernet}, {Verza}, {Viel}, {Vielzeuf}, {Vietri}, {Vink},
  {Viscasillas Vazquez}, {Wang}, {Weilbacher}, {Wendt}, {Wright}, {Ye},
  {Yeche}, {Yu}, {Zafar}, {Zibetti}, {Ziegler}, \&
  {Zinchenko}}]{2024arXiv240305398M}
{Mainieri}, V., {Anderson}, R.~I., {Brinchmann}, J., {et~al.} 2024, arXiv
  e-prints, arXiv:2403.05398

\bibitem[{{Mercer} {et~al.}(2005){Mercer}, {Clemens}, {Meade}, {Babler},
  {Indebetouw}, {Whitney}, {Watson}, {Wolfire}, {Wolff}, {Bania}, {Benjamin},
  {Cohen}, {Dickey}, {Jackson}, {Kobulnicky}, {Mathis}, {Stauffer}, {Stolovy},
  {Uzpen}, \& {Churchwell}}]{2005ApJ...635..560M}
{Mercer}, E.~P., {Clemens}, D.~P., {Meade}, M.~R., {et~al.} 2005, \apj, 635,
  560

\bibitem[{{Minniti} {et~al.}(2021{\natexlab{a}}){Minniti},
  {Fern{\'a}ndez-Trincado}, {Smith}, {Lucas}, {G{\'o}mez}, \&
  {Pullen}}]{2021A&A...648A..86M}
{Minniti}, D., {Fern{\'a}ndez-Trincado}, J.~G., {Smith}, L.~C., {et~al.}
  2021{\natexlab{a}}, \aap, 648, A86

\bibitem[{{Minniti} {et~al.}(2011){Minniti}, {Hempel}, {Toledo}, {Ivanov},
  {Alonso-Garc{\'\i}a}, {Saito}, {Catelan}, {Geisler}, {Jord{\'a}n},
  {Borissova}, {Zoccali}, {Kurtev}, {Carraro}, {Barbuy}, {Clari{\'a}},
  {Rejkuba}, {Emerson}, \& {Moni Bidin}}]{vvvcl001}
{Minniti}, D., {Hempel}, M., {Toledo}, I., {et~al.} 2011, \aap, 527, A81

\bibitem[{{Minniti} {et~al.}(2010){Minniti}, {Lucas}, {Emerson}, {Saito},
  {Hempel}, {Pietrukowicz}, {Ahumada}, {Alonso}, {Alonso-Garcia}, {Arias},
  {Bandyopadhyay}, {Barb{\'a}}, {Barbuy}, {Bedin}, {Bica}, {Borissova},
  {Bronfman}, {Carraro}, {Catelan}, {Clari{\'a}}, {Cross}, {de Grijs},
  {D{\'e}k{\'a}ny}, {Drew}, {Fari{\~n}a}, {Feinstein}, {Fern{\'a}ndez
  Laj{\'u}s}, {Gamen}, {Geisler}, {Gieren}, {Goldman}, {Gonzalez}, {Gunthardt},
  {Gurovich}, {Hambly}, {Irwin}, {Ivanov}, {Jord{\'a}n}, {Kerins}, {Kinemuchi},
  {Kurtev}, {L{\'o}pez-Corredoira}, {Maccarone}, {Masetti}, {Merlo},
  {Messineo}, {Mirabel}, {Monaco}, {Morelli}, {Padilla}, {Palma}, {Parisi},
  {Pignata}, {Rejkuba}, {Roman-Lopes}, {Sale}, {Schreiber}, {Schr{\"o}der},
  {Smith}, {}, {Soto}, {Tamura}, {Tappert}, {Thompson}, {Toledo}, {Zoccali}, \&
  {Pietrzynski}}]{2010NewA...15..433M}
{Minniti}, D., {Lucas}, P.~W., {Emerson}, J.~P., {et~al.} 2010, \na, 15, 433

\bibitem[{{Minniti} {et~al.}(2021{\natexlab{b}}){Minniti}, {Palma}, {Camargo},
  {Chijani-Saballa}, {Alonso-Garc{\'\i}a}, {Clari{\'a}}, {Dias}, {G{\'o}mez},
  {Pullen}, \& {Saito}}]{2021A&A...652A.129M}
{Minniti}, D., {Palma}, T., {Camargo}, D., {et~al.} 2021{\natexlab{b}}, \aap,
  652, A129

\bibitem[{{Moffat} {et~al.}(1991){Moffat}, {Shara}, \&
  {Potter}}]{1991AJ....102..642M}
{Moffat}, A. F.~J., {Shara}, M.~M., \& {Potter}, M. 1991, \aj, 102, 642

\bibitem[{{Morales} {et~al.}(2013){Morales}, {Wyrowski}, {Schuller}, \&
  {Menten}}]{2013A&A...560A..76M}
{Morales}, E. F.~E., {Wyrowski}, F., {Schuller}, F., \& {Menten}, K.~M. 2013,
  \aap, 560, A76

\bibitem[{{Muno} {et~al.}(2006){Muno}, {Bauer}, {Bandyopadhyay}, \&
  {Wang}}]{2006ApJS..165..173M}
{Muno}, M.~P., {Bauer}, F.~E., {Bandyopadhyay}, R.~M., \& {Wang}, Q.~D. 2006,
  \apjs, 165, 173

\bibitem[{{Nagata} {et~al.}(1993){Nagata}, {Hyland}, {Straw}, {Sato}, \&
  {Kawara}}]{1993ApJ...406..501N}
{Nagata}, T., {Hyland}, A.~R., {Straw}, S.~M., {Sato}, S., \& {Kawara}, K.
  1993, \apj, 406, 501

\bibitem[{{Nagata} {et~al.}(1995){Nagata}, {Woodward}, {Shure}, \&
  {Kobayashi}}]{1995AJ....109.1676N}
{Nagata}, T., {Woodward}, C.~E., {Shure}, M., \& {Kobayashi}, N. 1995, \aj,
  109, 1676

\bibitem[{{Nagata} {et~al.}(1990){Nagata}, {Woodward}, {Shure}, {Pipher}, \&
  {Okuda}}]{1990ApJ...351...83N}
{Nagata}, T., {Woodward}, C.~E., {Shure}, M., {Pipher}, J.~L., \& {Okuda}, H.
  1990, \apj, 351, 83

\bibitem[{{Nambiar} {et~al.}(2019){Nambiar}, {Das}, {Vig}, \&
  {Gorthi}}]{2019MNRAS.482.3789N}
{Nambiar}, S., {Das}, S., {Vig}, S., \& {Gorthi}, R. S.~S. 2019, \mnras, 482,
  3789

\bibitem[{Ochsenbein(1996)}]{10.26093/cds/vizier}
Ochsenbein, F. 1996, The VizieR database of astronomical catalogues

\bibitem[{{Ochsenbein} {et~al.}(2000){Ochsenbein}, {Bauer}, \&
  {Marcout}}]{vizier2000}
{Ochsenbein}, F., {Bauer}, P., \& {Marcout}, J. 2000, \aaps, 143, 23

\bibitem[{{Okuda} {et~al.}(1990){Okuda}, {Shibai}, {Nakagawa}, {Matsuhara},
  {Kobayashi}, {Kaifu}, {Nagata}, {Gatley}, \& {Geballe}}]{1990ApJ...351...89O}
{Okuda}, H., {Shibai}, H., {Nakagawa}, T., {et~al.} 1990, \apj, 351, 89

\bibitem[{{Paladini} {et~al.}(2023){Paladini}, {Zucker}, {Benjamin}, {Nataf},
  {Minniti}, {Zasowski}, {Peek}, {Carey}, {Allen}, {Alonso-Garcia}, {Alves},
  {Anders}, {Athanassoula}, {Beers}, {Bird}, {Bland-Hwathorn}, {Brown},
  {Buder}, {Casagrande}, {Casey}, {Cassisi}, {Catelan}, {Chary}, {Chene},
  {Ciardi}, {Comeron}, {Cohen}, {Dame}, {Drimmel}, {Fernandez Trincado},
  {Finkbeiner}, {Geisler}, {Gennaro}, {Goodman}, {Green}, {Hajdu}, {Henderson},
  {Hora}, {Ivanov}, {Kirkpatrick}, {Kobayashi}, {Kuhn}, {Kunder}, {Lu},
  {Lucas}, {Majaess}, {Megeath}, {Meisner}, {Molinari}, {Mroz}, {Ness},
  {Neumayer}, {Nogueras-Lara}, {Noriega-Crespo}, {Poleski}, {Rix}, {Rebull},
  {Reggiani}, {Rejkuba}, {Saito}, {Schoenrich}, {Saydjari}, {Schisano},
  {Schlafly}, {Schlaufman}, {Smith}, {Speagle}, {Wisz}, {Wyse}, \&
  {Zakamska}}]{2023arXiv230707642P}
{Paladini}, R., {Zucker}, C., {Benjamin}, R., {et~al.} 2023, arXiv e-prints,
  arXiv:2307.07642

\bibitem[{{Parsons} {et~al.}(2018){Parsons}, {Dempsey}, {Thomas}, {Berry},
  {Currie}, {Friberg}, {Wouterloot}, {Chrysostomou}, {Graves}, {Tilanus},
  {Bell}, \& {Rawlings}}]{2018ApJS..234...22P}
{Parsons}, H., {Dempsey}, J.~T., {Thomas}, H.~S., {et~al.} 2018, \apjs, 234, 22

\bibitem[{Parzen(1962)}]{10.1214/aoms/1177704472}
Parzen, E. 1962, The Annals of Mathematical Statistics, 33, 1065

\bibitem[{{Platais} {et~al.}(1998){Platais}, {Kozhurina-Platais}, \& {van
  Leeuwen}}]{1998AJ....116.2423P}
{Platais}, I., {Kozhurina-Platais}, V., \& {van Leeuwen}, F. 1998, \aj, 116,
  2423

\bibitem[{{Portegies Zwart} {et~al.}(2010){Portegies Zwart}, {McMillan}, \&
  {Gieles}}]{2010ARA&A..48..431P}
{Portegies Zwart}, S.~F., {McMillan}, S. L.~W., \& {Gieles}, M. 2010, \araa,
  48, 431

\bibitem[{{Preibisch} {et~al.}(2005){Preibisch}, {Kim}, {Favata}, {Feigelson},
  {Flaccomio}, {Getman}, {Micela}, {Sciortino}, {Stassun}, {Stelzer}, \&
  {Zinnecker}}]{2005ApJS..160..401P}
{Preibisch}, T., {Kim}, Y.-C., {Favata}, F., {et~al.} 2005, \apjs, 160, 401

\bibitem[{{Rathborne} {et~al.}(2016){Rathborne}, {Whitaker}, {Jackson},
  {Foster}, {Contreras}, {Stephens}, {Guzm{\'a}n}, {Longmore}, {Sanhueza},
  {Schuller}, {Wyrowski}, \& {Urquhart}}]{2016PASA...33...30R}
{Rathborne}, J.~M., {Whitaker}, J.~S., {Jackson}, J.~M., {et~al.} 2016, \pasa,
  33, e030

\bibitem[{{Rieke} \& {Lebofsky}(1985)}]{1985ApJ...288..618R}
{Rieke}, G.~H. \& {Lebofsky}, M.~J. 1985, \apj, 288, 618

\bibitem[{{Robitaille} {et~al.}(2008){Robitaille}, {Meade}, {Babler},
  {Whitney}, {Johnston}, {Indebetouw}, {Cohen}, {Povich}, {Sewilo}, {Benjamin},
  \& {Churchwell}}]{2008AJ....136.2413R}
{Robitaille}, T.~P., {Meade}, M.~R., {Babler}, B.~L., {et~al.} 2008, \aj, 136,
  2413

\bibitem[{{Ryu} \& {Lee}(2018)}]{2018ApJ...856..152R}
{Ryu}, J. \& {Lee}, M.~G. 2018, \apj, 856, 152

\bibitem[{{Sakai} {et~al.}(1996){Sakai}, {Madore}, \&
  {Freedman}}]{1996ApJ...461..713S}
{Sakai}, S., {Madore}, B.~F., \& {Freedman}, W.~L. 1996, \apj, 461, 713

\bibitem[{{Skrutskie} {et~al.}(2006){Skrutskie}, {Cutri}, {Stiening},
  {Weinberg}, {Schneider}, {Carpenter}, {Beichman}, {Capps}, {Chester},
  {Elias}, {Huchra}, {Liebert}, {Lonsdale}, {Monet}, {Price}, {Seitzer},
  {Jarrett}, {Kirkpatrick}, {Gizis}, {Howard}, {Evans}, {Fowler}, {Fullmer},
  {Hurt}, {Light}, {Kopan}, {Marsh}, {McCallon}, {Tam}, {Van Dyk}, \&
  {Wheelock}}]{2006AJ....131.1163S}
{Skrutskie}, M.~F., {Cutri}, R.~M., {Stiening}, R., {et~al.} 2006, \aj, 131,
  1163

\bibitem[{{Solin} {et~al.}(2012){Solin}, {Ukkonen}, \&
  {Haikala}}]{2012A&A...542A...3S}
{Solin}, O., {Ukkonen}, E., \& {Haikala}, L. 2012, \aap, 542, A3

\bibitem[{{Spitzer Science}(2009)}]{2009yCat.2293....0S}
{Spitzer Science}, C. 2009, VizieR Online Data Catalog, II/293

\bibitem[{{Stauffer} {et~al.}(2018){Stauffer}, {Helou}, {Benjamin}, {Marengo},
  {Kirkpatrick}, {Capak}, {Kasliwal}, {Bauer}, {Minniti}, {Bally}, {Lodieu},
  {Bowler}, {Zhang}, {Carey}, {Milam}, \& {Holler}}]{2018arXiv180600554S}
{Stauffer}, J., {Helou}, G., {Benjamin}, R.~A., {et~al.} 2018, arXiv e-prints,
  arXiv:1806.00554

\bibitem[{{Trumpler}(1930)}]{1930LicOB..14..154T}
{Trumpler}, R.~J. 1930, Lick Observatory Bulletin, 420, 154

\bibitem[{{Westerlund}(1961{\natexlab{a}})}]{1961PASP...73...51W}
{Westerlund}, B. 1961{\natexlab{a}}, \pasp, 73, 51

\bibitem[{{Westerlund}(1961{\natexlab{b}})}]{1961ArA.....2..419W}
{Westerlund}, B. 1961{\natexlab{b}}, Arkiv for Astronomi, 2, 419

\bibitem[{{Wright} {et~al.}(2010){Wright}, {Eisenhardt}, {Mainzer}, {Ressler},
  {Cutri}, {Jarrett}, {Kirkpatrick}, {Padgett}, {McMillan}, {Skrutskie},
  {Stanford}, {Cohen}, {Walker}, {Mather}, {Leisawitz}, {Gautier}, {McLean},
  {Benford}, {Lonsdale}, {Blain}, {Mendez}, {Irace}, {Duval}, {Liu}, {Royer},
  {Heinrichsen}, {Howard}, {Shannon}, {Kendall}, {Walsh}, {Larsen}, {Cardon},
  {Schick}, {Schwalm}, {Abid}, {Fabinsky}, {Naes}, \& {Tsai}}]{wise_descr}
{Wright}, E.~L., {Eisenhardt}, P. R.~M., {Mainzer}, A.~K., {et~al.} 2010, \aj,
  140, 1868

\bibitem[{{Zeidler} {et~al.}(2015){Zeidler}, {Sabbi}, {Nota}, {Grebel}, {Tosi},
  {Bonanos}, {Pasquali}, {Christian}, {de Mink}, \&
  {Ubeda}}]{2015AJ....150...78Z}
{Zeidler}, P., {Sabbi}, E., {Nota}, A., {et~al.} 2015, \aj, 150, 78

\bibitem[{{Zeidler} {et~al.}(2021){Zeidler}, {Sabbi}, {Nota}, \&
  {McLeod}}]{2021AJ....161..140Z}
{Zeidler}, P., {Sabbi}, E., {Nota}, A., \& {McLeod}, A.~F. 2021, \aj, 161, 140

\end{thebibliography}

\end{document}